\documentclass[preprint, noinfoline,addressatend,imslayout,a4paper,11pt]{imsart}

\usepackage{url}
\RequirePackage[colorlinks,citecolor=blue,urlcolor=blue]{hyperref}
\usepackage[authoryear]{natbib}

\usepackage{amsthm,amsmath,amsfonts,amssymb}
\usepackage{enumerate}
\usepackage{bm,accents}
\usepackage{graphicx}
\usepackage{color}

\startlocaldefs

\usepackage[plain,noend]{algorithm2e}

\makeatletter
\renewcommand{\algocf@captiontext}[2]{#1\algocf@typo. \AlCapFnt{}#2} 
\def\@algocf@capt@plain{top}
\renewcommand{\algocf@makecaption}[2]{%
	\addtolength{\hsize}{\algomargin}%
	\sbox\@tempboxa{\algocf@captiontext{#1}{#2}}%
	\ifdim\wd\@tempboxa >\hsize
	\hskip .5\algomargin%
	\parbox[t]{\hsize}{\algocf@captiontext{#1}{#2}}
	\else%
	\global\@minipagefalse%
	\hbox to\hsize{\box\@tempboxa}
	\fi%
	\addtolength{\hsize}{-\algomargin}%
}
\makeatother

\usepackage{longtable}

\usepackage{latexsym}
\usepackage{epsfig}
\usepackage{subfigure}
\usepackage{lscape}
\usepackage{latexsym}
\usepackage{epsfig}
\usepackage{graphicx}
\usepackage{amsmath,amsfonts,amssymb,bm}

\bibpunct{(}{)}{;}{a}{,}{,}
\def\spacingset#1{\renewcommand{\baselinestretch}%
	{#1}\small\normalsize} \spacingset{1.6}

\def\pr{\mbox{pr}}

\begin{document}
	\begin{frontmatter}
		
		
		
		
		
		\begin{aug}
			\title{$K$-sample omnibus non-proportional hazards tests based on right-censored data}
			\author{\fnms{Malka} \snm{Gorfine*}
				\ead[label=e1]{gorfinem@tauex.tau.ac.il}}
			\affiliation{Tel Aviv University, Israel}
			\printead{e1}
			
			\and
			\vspace{-0.2in}
			
			\author{\fnms{Matan} \snm{Schlesinger*}\ead[label=e2]{matan.schles@gmail.com} }
			\affiliation{Tel Aviv University, Israel}
			\printead{e2}
			
			\and
			
			\vspace{-0.2in}
			
			\author{\fnms{Li} \snm{Hsu}
				\ead[label=e3]{lih@fredhutch.org}}
			\affiliation{The Fred Hutchinson Cancer Research Center, Seattle USA}
			\printead{e3}
			
			\runtitle{$K$-sample omnibus non-proportional hazards tests}
			\runauthor{Gorfine, Schlesinger, Hsu}
			
		\end{aug}

		\begin{abstract}
			This work presents novel and powerful tests for comparing non-proportional hazard functions, based on sample-space partitions. Right censoring introduces two major difficulties which make the existing sample-space partition tests for uncensored data non-applicable: (i) the actual event times of censored observations are unknown; and (ii) the standard permutation procedure is invalid in case the censoring distributions of the groups are unequal. We overcome these two obstacles, introduce invariant tests, and prove their consistency. Extensive simulations reveal that under non-proportional alternatives, the proposed tests are often of higher power compared with existing popular tests for non-proportional hazards.  Efficient implementation of our tests is available in the R package KONPsurv, which can be freely downloaded from {https://github.com/matan-schles/KONPsurv}.
		\end{abstract}
		
		\begin{keyword} 
			\kwd{Consistent test} 
			\kwd{Crossing hazards} 
			\kwd{Non-parametric test} 
			\kwd{Non-proportional hazards} 
			\kwd{Permutation test} 
			\kwd{Right censoring} 
			\kwd{Sample-space partition}\\
			\kwd{*Both authors contributed equally to this work.}
		\end{keyword}
		
	\end{frontmatter}
	
\section{Introduction}

For the task of comparing survival distributions of two or more groups using censored data, the logrank test is the most popular choice. Its optimality properties under proportional-hazard functions are well known. Although the logrank test is asymptotically valid, it may not be powerful when the proportional hazards assumption does not hold. There are a variety of situations in which the hazard functions are of non-proportional shape. For example, a medical treatment might have adverse effects in the short run, yet effective in the long run, or a treatment may be short-term beneficial but gradually lose its effect with time. In such scenarios the hazard functions cross. In general, the longer the follow-up period is, the more likely it is for various non-proportional scenarios to develop \citep{yang2010improved}.

Other tests have been proposed that might be better choices for non-proportional hazards under the alternative. \citet{peto1972asymptotically} proposed a test which is similar to the logrank test, but more sensitive for differences in hazards at early survival times than at late ones. Pepe and Fleming \cite{pepe1989weighted,pepe1991weighted} suggested a weighted Kaplan--Meier (KM) test with a weight function consists of the geometric average of the two censoring survival-function estimators. \citet{yang2010improved} recently proposed another weighted logrank test whose weights are  obtained by fitting their model \citep{yang2005semiparametric}, which includes the proportional hazards and the proportional odds models as special cases. In contrast to the logrank and other related tests, the test of \citet{yang2010improved} uses adaptive weights. Under proportional hazards alternatives, this new adaptively-weighted logrank test is optimal. When the hazards are non-proportional, the adaptive weights typically lead to improvement in power over the logrank test. The test of \citet{yang2010improved}, referred  here as the Yang--Prentice test, is currently considered to be the leading one in terms of power, under a wide range of non-proportional hazards alternatives.
However, this test is applicable only for two-sample problems. Moreover, it is not invariant to group labeling. Exchanging the group labels between treatment and control would result in a different $p$-value. Thus, in applications with no clear link between the groups to treatment/control labeling, such as in testing the differences between females and males, the Yang--Prentice test in its current form is inappropriate. In Section~3.2 we suggest an invariant version of the Yang--Prentice test.

In the statistical literature of $K$-sample tests for non-censored data, there exist powerful consistent tests that are based on various sample-space partitions. These include the well-known Kolmogorv--Smirnov and Cramer--von Mises tests \citep{darling1957kolmogorov}, and the Anderson--Darling (AD) family of statistics \citep{pettitt1976two,scholz1987k}. In particular, \citet{thas2004extension} showed that the $K$-sample AD test is basically an average of Pearson statistics in $2 \times K$ contingency tables that are induced by observation-based partitions of the sample space into two subsets. They suggested an extension of the AD test, by considering  higher partitions, up to 4. \citet{heller2013consistent} proposed the HHG test, a sample-space partition-based non-parametric test for detecting associations between two random vectors of any dimension. When one of the random vectors is a categorical one-dimensional variable, the problem reduces to the $K$-sample problem with an observation-based partition of the sample space into two subsets, using three intervals. This specific partition is adopted in this work and will be described in detail. \citet{heller2016consistent} extended the work of \citet{thas2004extension} by considering sample-space partitions higher than 4, and studied test statistics that aggregate over all partitions by summation or maximization and also by aggregating over different sizes of partitions. They showed, by extensive simulation studies, that increasing the number of partitions can increase power under complex settings in which the density functions intersect 4 times or more.

In this work we present new powerful non-parametric and invariant tests for comparing two or more survival distributions using right-censored data. Our proposed methodology is demonstrated and applied using the specific sample-space partition of \citet{heller2013consistent}, which has been shown to be very powerful with 3 or less densities' intersections \citep{heller2016consistent}, under non-censored data. Right-censored data introduce two major difficulties: (i) the actual event times of censored observations are unknown; and (ii) the standard permutation procedure of label shuffling is invalid, in case the censoring distributions of the groups are unequal. We overcome these two obstacles and introduce novel consistent powerful tests for right-censored data.

\section{$K$-sample tests based on sample-space partition}
\subsection{Motivation and Notation}
Let $X$ be a one-dimensional non-negative random variable, $X\in \mathbb{R}^+$, and let $Y$ be a categorical variable indicating the group label. Under the $K$-sample hypotheses testing, the null hypothesis is  $H_0:F_1(x)=\ldots=F_K(x)$ for all $x\in \mathbb{R}^+$, and the alternative is $H_1:F_m(x)\neq F_k(x)$ for some $1\le m< k\le K$ and some $x\in \mathbb{R}^+$, where $F_k$ is the cumulative distribution function of group $k$, $k=1,\ldots,K$. We assume that the sample spaces on which these $K$ distributions are defined, coincide. 

$K$ random samples $\mathcal{A}_1,\ldots,\mathcal{A}_K$ are drawn from the respective distributions $F_1,\ldots, F_K$. Let $n_k$ be the total number of samples in group $k$, $k=1,\ldots,K$, and $n=\sum_{k=1}^{K}n_{k}$. Assume temporarily  no censoring, and consider a pair of observations $X_i \in \mathcal{A}_k$ and $X_j$. Then, in the spirit of the HHG test \citep{heller2013consistent}, we consider the following partition induced by the pair $(i,j)$:  
$A_{11}(i,j)$ is the number of observations from group $k$ that the distance between the value of $X$ and $X_i$ is less than or equal to $|X_i - X_j|$; $A_{12}(i,j)$ is the number of observations outside group $k$ that the distance to $X_i$ is less than or equal to $|X_i - X_j|$; $A_{21}(i,j)$ is the number of observations from group $k$ that the distance to $X_i$ is larger than $|X_i - X_j|$; and $A_{22}(i, j)$ is the number of observations outside group $k$ that the distance to $X_i$ is larger than $|X_i - X_j|$. 

\begin{figure}[!h]
	\caption{Sample-Space Partition}
	\centering
	\label{interval_illustration}
	\includegraphics[width=0.9\textwidth]{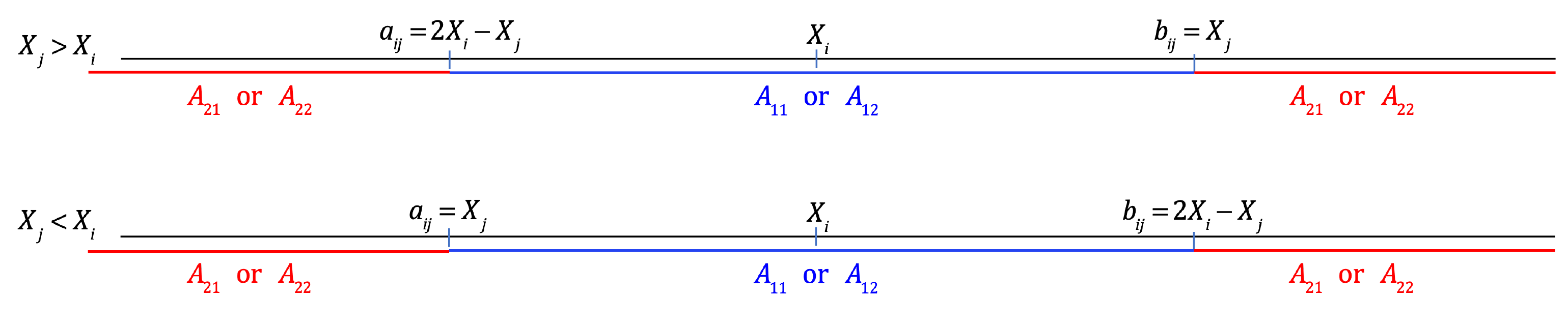}
\end{figure}

Interestingly, with no censoring, the $A_{lr}$'s, $l,r=1,2$, can be expressed through the empirical cumulative distribution functions. \textcolor{black}{For example, assume $X_i >X_j$ and $Y_i=Y_j=k$. Then, 
\begin{eqnarray}
A_{11}(i,j) &=& \sum_{r=1,r\neq i,j}^n I\{|X_i-X_r| \leq X_i-X_j \}I\{Y_r=k\} \nonumber \\
&=& \sum_{r=1,r\neq i,j}^n I \{ X_j \leq X_r \leq 2X_i - X_j \}I\{Y_r=k\} \nonumber \\
&=&  n_k\{\widehat{F}_k(2X_i - X_j) - \widehat{F}_k(X_{j}^-) \} - 2  \nonumber \, ,
\end{eqnarray}
where $\widehat{F}_k(x)= n_k^{-1} \sum_{X_l\in \mathcal{A}_k} I (X_l \le x) $ and $\widehat{F}_k(x^-)= n_k^{-1}\sum_{X_l\in \mathcal{A}_k} I (X_l<x)$. The $-2$ above stands for excluding observations $i$ and $j$. In general, all observations with a distances to $X_i$ that is less than or equal to $|X_i - X_j|$ lie inside the interval $[a_{ij},b_{ij}]$, where $a_{ij}=\min(X_j,2X_i-X_j)$ and $b_{ij}=\max(X_j,2X_i-X_j)$, as illustrated in Figure \ref{interval_illustration}. }

In general, for $Y_i=k$, we get 
\begin{eqnarray}
A_{11}(i,j) &= & n_k\{\widehat{F}_k(b_{ij}) - \widehat{F}_k(a_{ij}^-) \} -1 - I(Y_i=Y_j)\, , \nonumber \\
A_{12}(i,j) &=& \sum_{m=1,m\neq k}^{K}n_m\big\{\widehat{F}_{m}(b_{ij})  - \widehat{F}_{m} (a_{ij}^-) \big\} -I(Y_i \neq Y_j) \, , \nonumber \\
A_{21}(i,j) &=& n_k - A_{11}(i,j) - 1 - I(Y_i=Y_j) \, , \nonumber \\
A_{22}(i,j) &=& \sum_{m=1,m\neq k}^{K}n_m - A_{12}(i,j) - I(Y_i \neq Y_j) \, . \nonumber
\end{eqnarray}
 For each pair $(i,j)$, a $2 \times 2$ contingency table can be constructed with $A_{lr}(i,j)$  as the entry of cell $lr$, $l,r=1,2$, and a total sum of $n-2$. \textcolor{black}{Under the null hypothesis of equal distributions, the probability of belonging to cell $lr$ equals the product of the marginal probabilities of the $l$th row and
 the $r$th column.} Therefore, a summary statistic of a such contingency table can be based on either the Pearson chi-squared test statistic, or the log-likelihood ratio statistic.   
 {Since it is unknown which pair $(i,j)$ yields the best sample-space partition that provides the largest summary statistic, the final test statistic is defined as a sum over all possible $n(n-1)$ partitions induced by the data.} A permutation-based $p$-value can be calculated, under random permutations of the group labels.

To introduce the right-censored data, let $C \in \mathbb{R}^+$ be a non-negative random variable indicating the censoring time. Assume that $X$ and $C$ are conditionally independent given $Y$. Define $T$ to be the observed time, namely $T=\min(X,C)$ and let $\Delta=I(X\le C)$. Hence, the observed data consist of $K$
random samples that can be summarized by  $(T_i, \Delta_i, Y_i),\ i = 1,\ldots,n$. Note that the different groups may have different censoring distributions. With right-censored data, our proposed test requires special care in evaluating the $A_{r,l}$'s, $r,l = 1,2$, by replacing the empirical distribution functions by their respective Kaplan--Meier (KM) estimators, as well as when applying a permutation test with unequal censoring distributions. Both issues are described in details below.

\subsection{The Test Statistic} 
Let $\widetilde{F}_k$ be the KM estimator of the cumulative distribution function using all observations of group $k$. The KM estimator is defined only up to (and including) the last observed failure time. Define $\gamma_k$ to be the maximum time in which $\widetilde{F}_k$ can be used for the test statistic. If the largest observed time is a failure time, $\widetilde{F}_k(t)$ is known for the entire range of $t$. In this case, we define $\gamma_k$ to be the maximum possible value of $t$ required for the test, $2 \max_{i=1,\ldots,n}\{T_i\} - \min_{i=1,\ldots,n}\{T_i\}$. However, in case of censoring after the largest failure time, the KM estimator beyond the largest observed failure time is undefined. We thus define the maximum time  $\gamma_k$ to be the largest observed failure time.  Namely,
$$\gamma_k = \left\{
        \begin{array}{ll}
            2\displaystyle\max_{i=1,\ldots,n}\{T_i\} - \displaystyle\min_{i=1,\ldots,n}\{T_i\} & \quad 
\mbox{if} \,\,\,\,            
            \displaystyle\max_{i=1,\ldots,n} \{T_i \Delta_i I(Y_i=k) \}=\displaystyle\max_{i=1,\ldots,n} \{T_i I(Y_i=k) \}  \\
            \displaystyle\max_{i=1,\ldots,n} \{T_i \Delta_i I(Y_i=k)\} & \quad 
\mbox{if} \,\,\,\,               
            \displaystyle\max_{i=1,\ldots,n} \{T_i \Delta_i I(Y_i=k) \} \neq \displaystyle\max_{i=1,\ldots,n} \{T_i I(Y_i=k) \}
        \end{array}
    \right. $$
Define $\gamma_{-k}$ to be the maximum time in which at least one of the other KM estimators,  $\widetilde{F}_m$, $m=1,\ldots,K, m\neq k$, can be used for the test statistic, namely $\gamma_{-k}=\max_{m\neq k}\{\gamma_m\}$. Define $\tau_k$ to be the maximum time point the KM estimator is defined in group $k$ and in at least one more group, namely $\tau_k=\min\{\gamma_k,\gamma_{-k}\}$.
Then, for each pair of observed failure times $T_i\in \mathcal{A}_k$ and $T_j$ such that  $j\neq i,\ \Delta_i= \Delta_j=1$ and $b_{ij} \le \tau_k$, a $2\times2$ contingency table is constructed. The following $A_{lr}^*(i,j),\ l,r=1,2$, are the corresponding expressions of $A_{lr}(i,j)$ obtained by replacing $\widehat{F}$ by $\widetilde{F}$, the KM estimators. In case $\gamma_m<b_{ij}$, the observations of group $m$ will not be included in the contingency table induced by the pair $(i,j)$. Namely, for $j\neq i,\ \Delta_i= \Delta_j=1, b_{ij} \le \tau_k$ and $Y_i=k$,
\begin{eqnarray}
A_{11}^*(i,j)&=& n_k\{\widetilde{F}_k (b_{ij}) - \widetilde{F}_k(a_{ij}^-) \} -1-I(Y_i=Y_j)\, ,\nonumber \\
A_{12}^*(i,j)&=& \sum_{m=1,m\neq k}^{K}n_m\big\{\widetilde{F}_{m}(b_{ij})  - \widetilde{F}_{m} (a_{ij}^-) \big\}I(\gamma_m\ge b_{ij})\ -I(Y_i\neq Y_j)\, , \nonumber \\
A_{21}^*(i,j) &=& n_k - A_{11}^*(i,j)  -1-I(Y_i=Y_j)\, , \nonumber  \\
A_{22}^*(i,j) &=& \sum_{m=1,m\neq k}^{K}n_mI(\gamma_m\ge b_{ij}) - A_{12}^*(i,j)-I(Y_i\neq Y_j) \, . \nonumber
\end{eqnarray}
Only pairs of observed failure times are used for the sample-space partitioning (i.e., $\Delta_i=\Delta_j=1$), while censored observations contribute in $\widetilde{F}_k$, $k=1,\ldots,K$. Denote by $n(i,j)$ the number of observations in all the groups included in the contingency table induced by $(i,j)$, namely $n(i,j)=\sum_{k=1}^{K}n_kI(\gamma_k\ge b_{ij})$. Let $S_P$ and $S_{LR}$ be the summary statistic of each contingency table, based on Pearson chi-squared test statistic
$$
S_P(i,j)=\frac{\{n(i,j)-2\}\{A^*_{12}(i,j)A^*_{21}(i,j)-A^*_{11}(i,j)A^*_{22}(i,j)\}^2}{A^*_{1\cdot}(i,j)A^*_{2\cdot}(i,j)A^*_{\cdot 1}(i,j)A^*_{\cdot 2}(i,j)} \, ,
$$
and the log-likelihood ratio statistic,
$$
S_{LR}(i,j)=2\sum_{m=1,2}\sum_{k=1,2}A^*_{mk}(i,j)\log \bigg[\frac{\{n(i,j)-2\}A^*_{kl}(i,j)}{A^*_{m\cdot}(i,j )A^*_{\cdot k}(i,j)} \bigg] \, ,
$$
respectively, where $A^*_{.k}(i,j)=  \sum_{m=1,2}A^*_{mk}(i,j)$ and $A^*_{m.}(i,j)=\sum_{k=1,2}A^*_{mk}(i,j)$.
In case of at least one zero margin in the contingency table, $S_P(i, j)=0$ and
$S_{LR}(i, j)=0$. Denote by 
$$
N=\sum_{k=1}^{K}\sum_{i=1, T_i \in \mathcal{A}_k}^n\sum_{j=1,j\neq i}^{n} \Delta_i \Delta_j I(T_j\le \tau_k) 
I(b_{ij} \le \tau_k)
$$ 
the total number of tables constructed from the data. Then, our proposed sample-space partition test statistic for equality of $K$ distributions based on right-censored data is defined by 
$$
Q = \frac{1}{N}\sum_{k=1}^{K}\sum_{i=1, T_i \in \mathcal{A}_k}^n\sum_{j=1,j\neq i}^{n} S(i,j) \Delta_i \Delta_j I(b_{ij} \le \tau_k)\, .
$$
where $S(i,j)$ is either the test statistic $S_P(i,j)$ or $S_{LR}(i,j)$.
In the case of no right censoring, the number of tables is solely determined by the number of observations. In right-censored data, the number of tables is random and determined also by the actual observed values due to the restrictions $\Delta_i= \Delta_j=1$ and $b_{ij} \le \tau_k$. This issue is of high importance for the permutation stage of the test, as elaborated in the next subsection.

\subsection{The Permutation Procedure}
\label{Test_stat_section}
Allegedly, a permuted test can be done based on random permutations of the group labels. However, if the censoring distributions of the $K$ groups are different, such a permutation test is invalid, since a significant result can be yielded under the null due to differences in the censoring distributions. In order to generate random permutations that are independent of $Y$, we adopt the imputation approach suggested by \citet{wang2010testing}.

The main idea consists of randomly permuting the group labels, while for each observation assigned to a group different from the original one, a censoring time is imputed from the  censoring distribution of the new assigned group. If the observation was originally censored, a survival time is also imputed, from the null survival distribution. Let $Y_1^{p},...,Y_n^{p}$ be a random permutation of the group labels. Define $(T_i^p, \Delta_i^p),\ i = 1, \ldots,n, $ by 
$$
(T_i^p, \Delta_i^p) = \left\{
        \begin{array}{ll}
            (T_i,\Delta_i) & \quad \mbox{if} \,\,\,\, Y_i^p=Y_i \\
            (\widetilde{T}_i,\widetilde{\Delta}_i) & \quad \mbox{if} \,\,\,\, Y_i^p \neq Y_i
        \end{array} \,\,\,\, ,
    \right.
$$
where $\widetilde{T}_i=\min(\widetilde{X}_i,\widetilde{C}_i)$, $\widetilde{\Delta}_i=I(\widetilde{X}_i \le \widetilde{C}_i)$. Furthermore, $\widetilde{C}_i$ is sampled from the estimated censoring distribution of group $Y_i^p$, based on the KM estimator of the censoring distribution of group $Y_i^p$, by reversing the roles of event and censoring. The KM estimator, denoted by $\widehat{G}_{C,Y_i^p}$, is defined up to the largest observed censoring value of that group. Then, each observed censoring time is sampled with probability equals to the jump size of the respective KM estimator. In case $\widehat{G}_{C,Y_i^p}$ is incomplete, i.e., $\widehat{G}_{C,Y_i^p}\big[\max_{j=1,\ldots,n} \{T_j(1-\Delta_j) I(Y_j=Y_i^p)\}\big]>0$, the largest observed time, $\max_{j=1,\ldots,n} \{T_j (1-\Delta_j)I(Y_j=Y_i^p)\}$, is sampled with probability $\widehat{G}_{C,Y_i^p}\big[\max_{j=1,\ldots,n} \{T_j(1-\Delta_j) I(Y_j=Y_i^p)\}\big]$. $\widetilde{X}_i$ is defined by
$$
\widetilde{X}_i = \left\{
        \begin{array}{ll}
            X_i & \quad \mbox{if} \,\,\,\, \Delta_i=1 \\
            X_i^* & \quad \mbox{if} \,\,\,\, \Delta_i=0
        \end{array} \,\,\,\, ,
    \right.
$$
where $X_i^{*}$ is sampled from an estimator of $\pr (X_i>x|X_i>T_i)$, the conditional distribution of $X$ under the null hypothesis. In practice, $\pr (X_i>x|X_i>T_i)$ is replaced by its KM estimator, using all observations from all groups that their observed time is larger than $T_i$. Denote this KM estimator by $\widehat{S}_{cond,T_i}$. The sampling based on $\widehat{S}_{cond,T_i}$ is done in the same fashion as in the above censoring sampling, but in case of incomplete distribution, the value $\max_{i=1,\ldots,n} (T_i \Delta_i) +\varepsilon$ ($\varepsilon $ is any positive number) is sampled with probability $\widehat{S}_{cond,T_i}\{\max_{i=1,\ldots,n} (T_i\Delta_i)\}$, and its respective event indicator is set to be $\Delta_i^p=0$, as there is no empirical evidence for the potential failure times beyond $\max_{i=1,\ldots,n}\{T_i \Delta_i\}$. 

When performing a permutation test, the reported $p$-value can be viewed as an approximation of the true $p$-value, based on all possible permutations. In the above imputation-based permutation procedure, additional variability in a $p$-value is expected due to random imputations. To reduce this variability, multiple imputations can be used, such that for each random imputation, $B$ permutations are generated. Assume $M$ imputations are used. Then the $p$-value is defined as the fraction of the test statistics among the $MB$ test statistics that are at least as large as the observed test statistic $Q$.

In the following theorem it is argued that our proposed tests are consistent against all alternatives. The proof is presented in details in the Supplementary Materials. \\

\textbf{Theorem:}
Let $X$ be a positive failure time random variable, either continuous or discrete, and $Y$ be a categorical random variable with $K$ categories. Let $\pi_k= \lim_{n\to\infty} n_k/n$, $k=1,\ldots,K$. Assume there are at least two cumulative distribution functions $F_g(x)$ and $F_m(x),\ g,m \in \{1,\ldots,K\}$, such that $F_g(x_0)\neq F_m(x_0)$ for some $x_0 \in \mathbb{R}^+$, $\pi_g>0,\ \pi_m>0$, and the conditional censoring distributions are such that  $\pr(C > x_0|Y=g)>0$ and $\pr(C > x_0|Y = m)>0$. Then, the imputation-based permutation test with the test statistic $Q$ is consistent, namely, the power of the test increases to $1$ as $n\to \infty$.

\subsection{Computation Time}
Table \ref{running_time_table} provides the run time of the proposed tests of one dataset, $K=2$, under the null hypothesis, one imputation, and 1000 permutations, for different total sample sizes $n$ and $n_1=n_2=n/2$. These results were generated by a 6 years old Intel i7-3770 CPU 3.4 GHz, without paralleling with the different cores of the computer. The first two rows are for identical censoring distributions, and the last two are for different censoring distributions. Evidently, even with $n=1000$ observations and low censoring rates, the run time on such a simple computer is not longer than 3.5 minutes. We expect that an increase of the number of imputations and permutations will increase the run time in a linear fashion. 

\section{Simulation study}
\label{Simulation_study}

\subsection{Simulation Design}
\label{Simulation_design_section}
An extensive numerical study has performed to systematically examine the behavior of our proposed $K$-sample omnibus non-proportional hazards (KONP) tests under a wide range of alternatives, various sample sizes, and a wide range of censoring distributions, including unequal censoring distributions. The main part of the simulation study was dedicated to the popular 2-sample setting, but settings of $K=3,4,5$ were considered as well.

As competitors under the 2-sample setting, the following tests were included: the logrank test; Peto--Peto weighted logrank test~\citep{peto1972asymptotically} {that uses a weight function that is very close to the pooled KM estimator}; Pepe--Fleming weighted KM test~\citep{pepe1989weighted} {with geometric mean of the two KM censoring-distribution estimators as a weight function}; and Yang--Prentice test, {an adaptive weighted logrank test where the adaptive weights utilize the hazard ratio obtained by fitting the model of Yang and Prentice~\cite{yang2005semiparametric}}. The tests of~\citet{uno2015versatile}  are invalid under unequal censoring distributions (as demonstrated below), and thus are not included in the following power comparisons.

Table \ref{unequal_cen} (main text) and tables \ref{scenario_description} and \ref{prop_design} in Appendix, provide a comprehensive summary of the 17 non-proportional hazards scenarios and 7 proportional or close to proportional hazards functions, that were studied. For each scenario, the failure and censoring distributions are explicitly provided, and the survival functions of the two groups are plotted. A reference is provided indicating the source of each setting. In short, Scenario A shows differences at mid time points, but similarity in early and late times. Scenarios B--D show differences in early times. Scenario E is of equal survival functions at early times and of proportional hazards at mid and late times. Scenarios F and G are with crossing hazards. Scenario H is of a U-shape hazards ratio. Scenarios I, J and K are with crossing hazards, based on the following hazards-ratio model of \citet{yang2005semiparametric}:
\begin{equation}\label{hr_yp}
\mbox{HR}(t)=\frac{\lambda_2(t)}{\lambda_1(t)}=\frac{\theta_1 \theta_2}{\theta_1 +(\theta_2-\theta_1)S_1(t)}\, , \,\,\,\,  0<t<\tau_0 \,, \,\,\,\, \theta_1,\theta_2>0 \,  ,
\end{equation}
where $\tau_0=\sup \{t:S_1(t)>0 \}$, $\lambda_2$ and $\lambda_1$ are the hazard functions of the two groups, and $S_1$ is the survival function of group $1$. Under Model (\ref{hr_yp}), $\theta_1 = \lim_{t\to 0} \mbox{HR}(t)$ and 
$\theta_2= \lim_{t\to \tau_0} \mbox{HR}(t)$. Also it is assumed that for a continuous function $S_1$, $\mbox{HR}(t)$ is a strong monotone function of $t$, i.e. 
$\mbox{sign} \{ d \mbox{HR}(t)/dt \}$ is the same for all $t \in (0, \tau_0)$. 
Scenarios I-1, I-2 and I-3 are with crossing hazards under Model (\ref{hr_yp}). In I-1 the hazard functions cross earlier compared to I-2 and I-3.
Scenarios J-1, J-2 and J-3 are with crossing hazards in which the strong monotonicity  assumption is violated. In J-1 and J-2 the hazards are piece-wise proportional, and the hazard functions cross in mid time points. In Scenario J-3 the hazard ratio is a continuous function of 
$t$, and the hazards cross at a late time point. Under Scenarios K-1, K-2 and K-3 Model (\ref{hr_yp}) is violated, but the strong monotonicity  assumption of $\mbox{HR}(t)$ holds. In Scenarios K-1 and K-2 the hazards cross at early-mid times, and in Scenario K-3 at mid times.

For each scenario described above, four different censoring distributions were considered, two with equal and two with unequal censoring distributions. 
Under equal censoring distributions, the censoring distributions were taken to be similar to the corresponding referenced paper. Exponential distributions were used for all other scenarios, with approximately  $25\%$ or $50\%$ censoring rates.
Under unequal censoring distributions, the censoring distributions of \citet{wang2010testing} were used (Table \ref{unequal_cen}). The specific values of $(a,b,\theta_1,\theta_2)$ are provided in tables \ref{scenario_description}-\ref{prop_design}. Under small differences, the censoring rates among the two groups are about $40\%$ and $55\%$, where $27\%$ and $55\%$ are the respective values under substantial differences. The various censoring settings are such that the power of a specific test under specific scenario is not necessarily  increasing as the censoring rate decreases. 

Each of the configurations was studied with $n=100,200,300$ or 400, $n_1=n_2$, and performances are summarized based on 2000 replications.

A smaller simulation study was done for $K>2$. As competitors, the logrank and Peto-Peto tests were included. For the null scenario $K=3,4,5$ were studied, and under Scenarios D and J-2, $K=3$ was examined. Various sample sizes and a wide range of censoring distributions were considered. A detailed description of these scenarios can be found in the Supplementary Materials.

\subsection{The test of \citet{yang2010improved}}
\label{YP_inv}
Since the Yang--Prentice test is the strongest competitor in terms of power, for the $2$-sample setting, we highlight some of its properties. The Yang--Prentice test is based on Model (\ref{hr_yp}), where the indices 1 and 2 indicate the control and treatment groups, respectively. Since this model is asymmetric in terms of $F_1$ and $F_2$, the test is not group-label invariant. By exchanging the group labels between `treatment' and `control', a different $p$-value would be provided. This property is unique to this test, and all other tests considered in this work are invariant to group labeling. Consequently, in applications with no clear link between the two groups to treatment/control status (e.g., comparing  females versus males, or young versus old), it is unclear how the Yang--Prentice test should be applied.

In order to make the Yang--Prentice test invariant, we apply a permutation test based on the minimum $p$-value of the two labeling options. Specifically, let $PV_1$ and $PV_2$ be the $p$-values of the Yang--Prentice tests based on $(X_i, \Delta_i, Y_i),\ i = 1, \ldots,n$, 
and $(X_i, \Delta_i, \widetilde{Y}_i),\ i = 1, ...,n$, respectively, where $Y_i\in\{1,2\}$ and 
$\widetilde{Y}_i=I(Y_i=2)+2I(Y_i=1)$. Then, the Yang--Prentice invariant test statistic is defined by $Q_{YP}=\min \{PV_1,PV_2\}$.  The $p$-value of the imputation-based permutation test, based on the statistic $Q_{YP}$, is the fraction of replicates of $Q_{YP}$ under random permutations of the data, as described in Section 2.3, that are at least as small as the observed test statistic.
Since our work mainly focuses on invariant tests, we report and discuss in the main text the performance of the Yang--Prentice invariant test. The  results of the original Yang-Prentice test, with the two possible options of labeling for treatment and control, can be found in the Supplementary Materials.

The original Yang--Prentice test (implemented in the R package YPmodel) uses the asymptotic distribution of the test statistic under the null hypothesis. Often, its empirical size is greater than the nominal level, especially under small sample size (see the Supplementary Material). In a recent work  of \citet{yang2019interim}, which deals with interim monitoring using adaptive weighted log-rank test, the author suggests using the re-sampling method of \citet{lin1993checking} instead of the original asymptotic approach \citep{yang2010improved}, for improving the type I error rate. The method of \citet{lin1993checking} is also based on asymptotic results. Alternatively, the imputation approach of \citet{wang2010testing} works very well in very small sample sizes, and it does not require asymptotic distribution. To make the comparison between the methods consistently, we use the imputation approach for both, our proposed method and the invariant version of Yang-Prentice test.

\subsection{Simulation Results}
\label{Simulation_results_section}
\begin{figure}
	\caption{Empirical Power Under the Null: top - equal censoring rates, bottom - unequal censoring rates}
	\centering
	\includegraphics[width=0.55\textwidth]{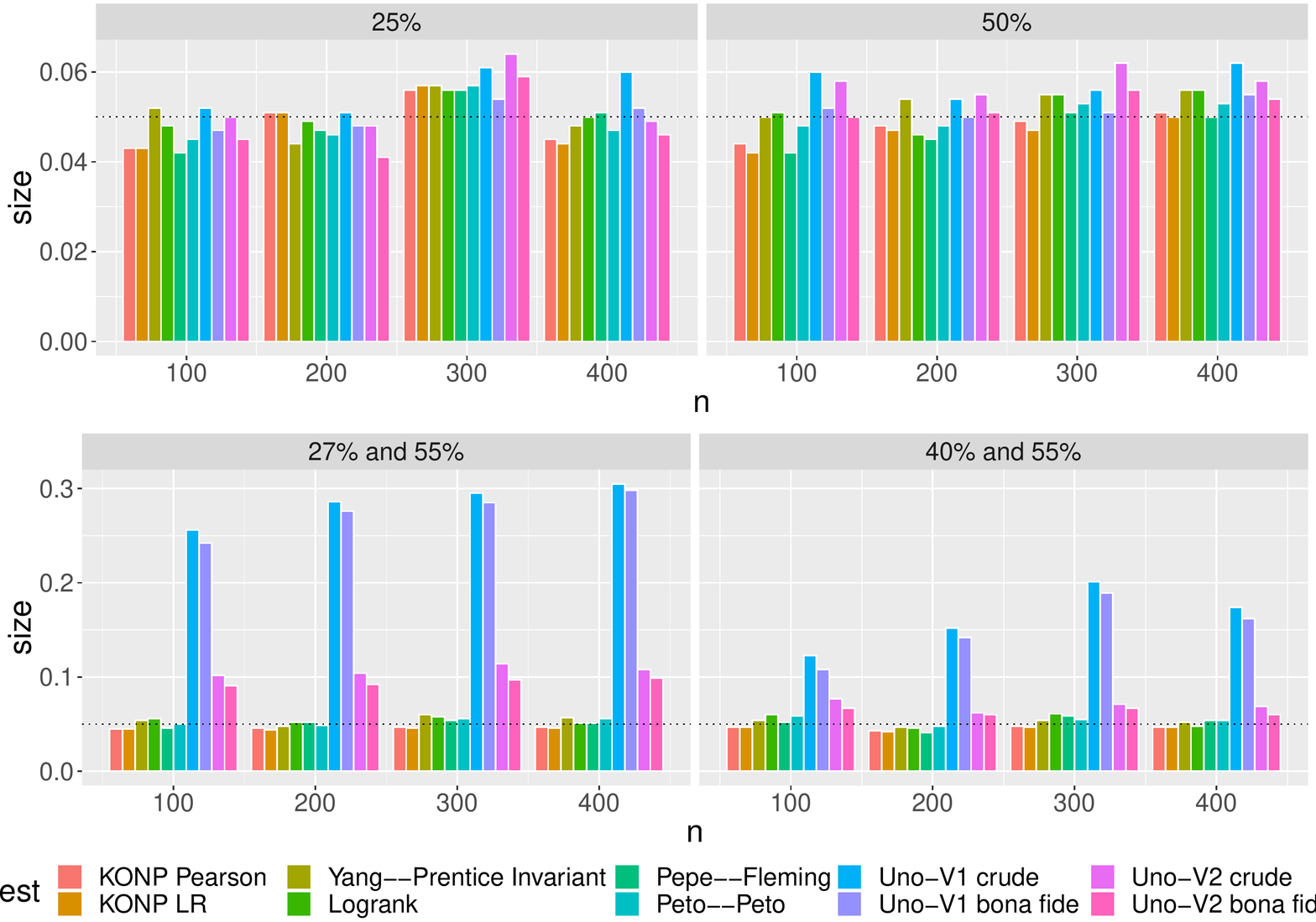}
	\label{Null_equal_censoring}
\end{figure}

Figure \ref{Null_equal_censoring} provides the empirical power of the tests under the null hypothesis, with equal and unequal censoring distributions. Evidently, under equal censoring distributions all the tests are valid, as the empirical size of the tests are reasonably close to the nominal value $0.05$. On the other hand, under the null hypothesis and unequal censoring distributions, the empirical sizes of Uno et al. tests are much higher than the nominal value 0.05. For example, under a sample size of $n=400$, and censoring rates of approximately $27\%$ and $55\%$, the empirical size of Uno et al. $V_2$ bona fide test is $0.099$; all the other Uno et al. tests' sizes are even higher. The empirical sizes of all the other tests are reasonably close to their nominal value. Thus, Uno's tests will not be considered in the rest of this simulation study.

\begin{figure}[!h]
	\caption{Simulation Results of non-proportional hazards settings considered by others:  {Setting A shows differences at mid time points, but similarity in early and late times. B--D show differences in early times. E is of equal survival functions at early times and of proportional hazards at mid and late times. Scenarios F and G are with crossing hazards under model (\ref{hr_yp}). Scenario H is of a U-shape hazards ratio.}}
	\centering
	\label{ref_scen_fig}
	\includegraphics[width=1.00\textwidth]{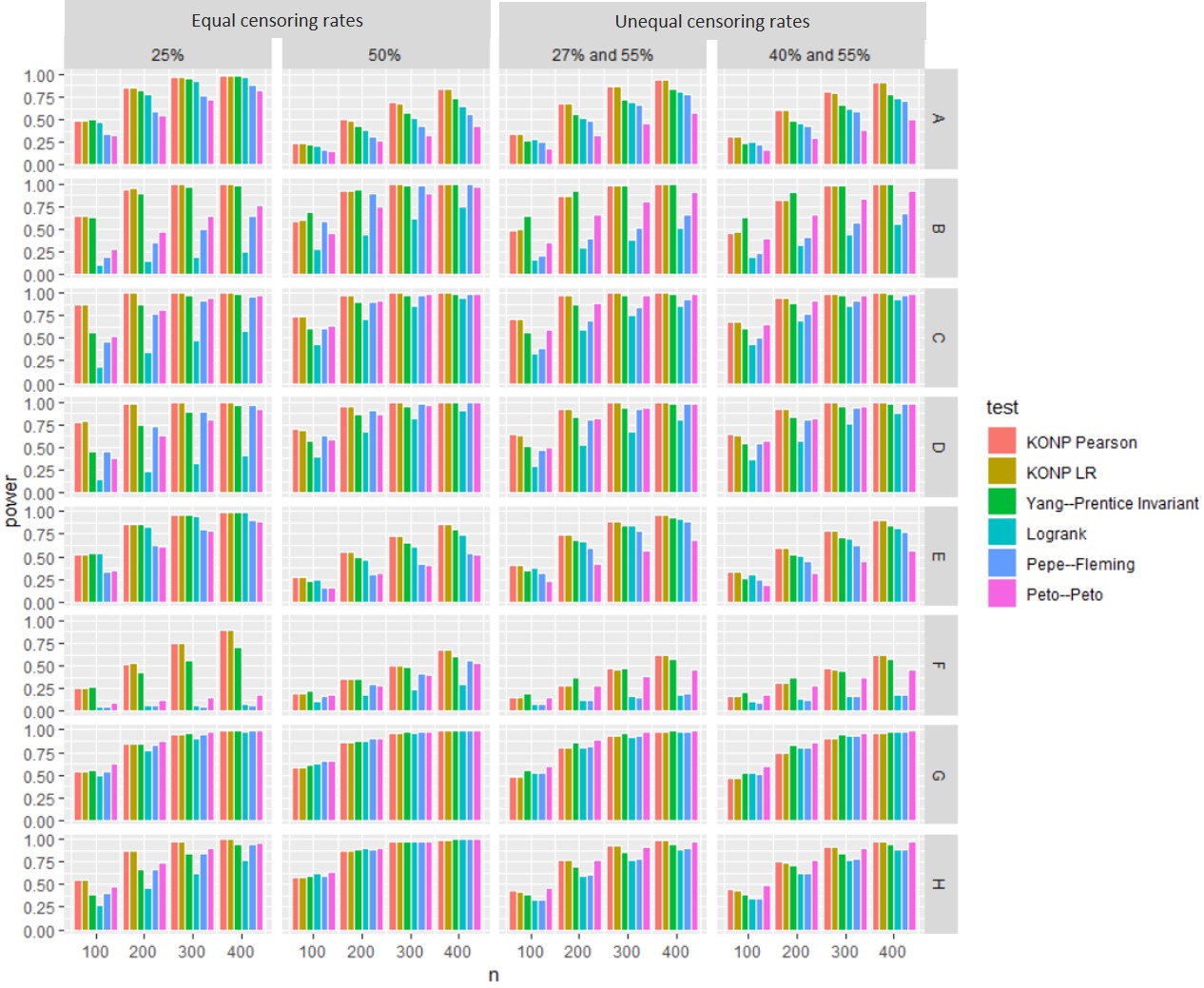}
\end{figure}

\begin{figure}[!h]
	\caption{Simulation results of additional settings of non-proportional hazards: {Scenarios I-1, I-2 and I-3 are with crossing hazards under Model (\ref{hr_yp}). In I-1 the hazard functions cross earlier compared to I-2 and I-3.
		Scenarios J-1, J-2 and J-3 are with crossing hazards in which the strong monotonicity  assumption is violated. In J-1 and J-2 the hazards are piece-wise proportional, and the hazard functions cross in mid time points. In Scenario J-3 the hazard ratio is a continuous function of 
		$t$, and the hazards cross at a late time point. Under Scenarios K-1, K-2 and K-3 Model (\ref{hr_yp}) is violated, but the strong monotonicity  assumption of $\mbox{HR}(t)$ holds. In Scenarios K-1 and K-2 the hazards cross at early-mid times, and in Scenario K-3 at mid times.}}
	\centering
	\label{our_scen_fig}
	\includegraphics[width=1.00\textwidth]{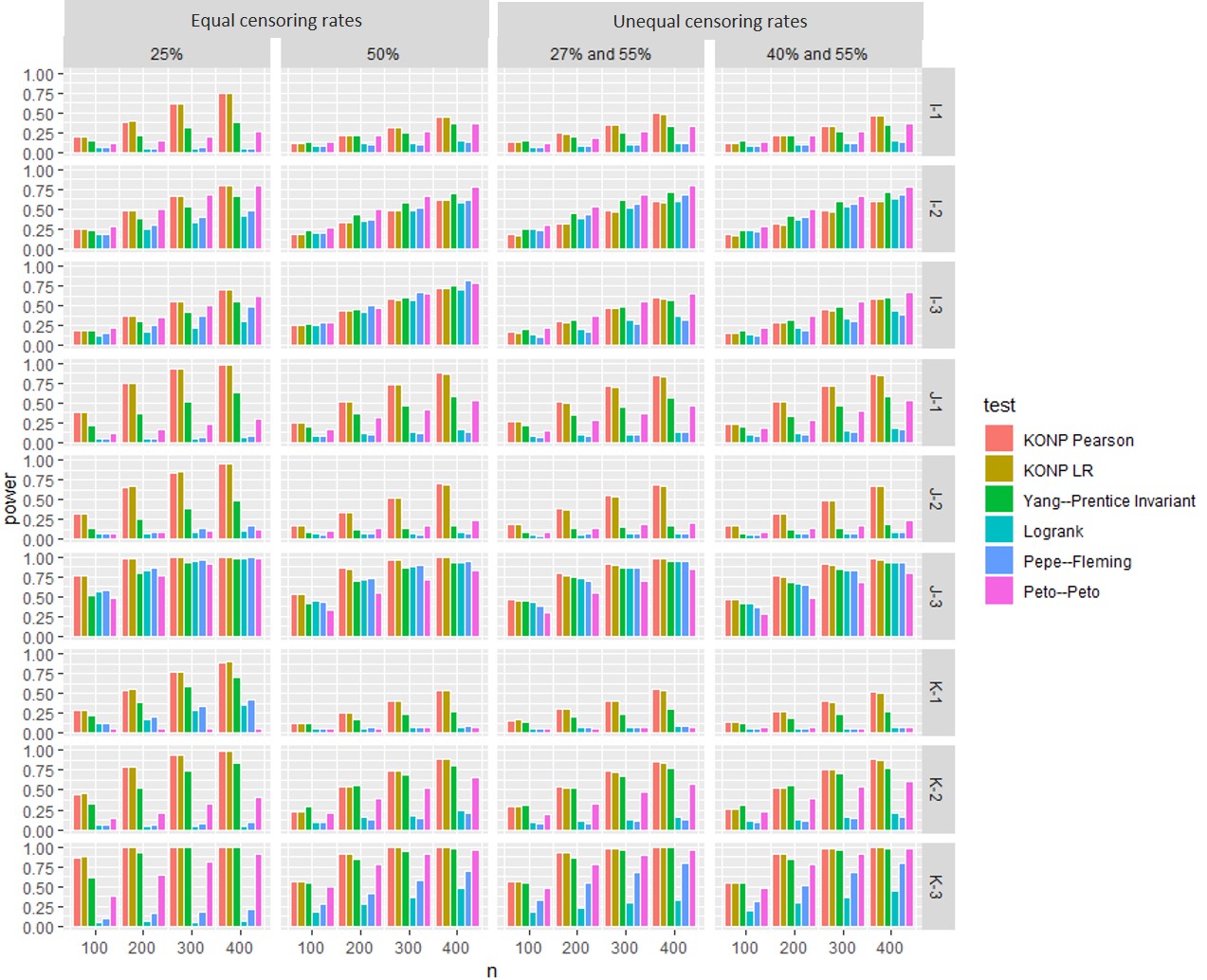}
\end{figure}

Figure  \ref{ref_scen_fig} summarizes the empirical power of the tests under settings generated by others, while  Figure \ref{our_scen_fig} is based on scenarios generated by us.  As expected, the power of each test increases with the sample size.  In some scenarios the power of the tests increases as the censoring rate increases, since in these settings the non-censored data are centered mainly at the parts of the hazards which are closer to proportionality and the censored data are mainly located at the non-proportionality area of the hazards. For example, in Scenario I-3 with $n=400$, as the censoring rate increases from about $25\%$ to about $50\%$, the power of Peto--Peto increases by $0.159$, Yang--Prentice by $0.2$, Pepe--Fleming by $0.218$, and logrank by $0.4$. The power increase in our KONP tests is much smaller, 0.012 and 0.009.

{Evidently, our KONP tests are often more powerful compared to all other tests. These includes
scenarios A, C, D, E, F, J-1, J-2, J-3, I-1, K-1, K-2, K-3.   
The superiority of KONP over Yang-Prentice under F and I-1 is surprising, since these scenarios follow their Model (\ref{hr_yp}). For example, with $n=400$ and $25\%$ censoring rate, the empirical power is about $90\%$ for KONP tests and only $70\%$ for Yang--Prentice. In Scenario G (close to proportional hazards), which also follows Model (\ref{hr_yp}), the results of Yang-Prentice and logrank are similar and often slightly better than KONP tests. In scenarios J-1 and J-2 our tests are substantially more powerful than all the competitors. For example, in Scenario J-2, $n=400$, and $25\%$ censoring rate, KONP tests  with about $95\%$ power, while the second most powerful test is Yang--Prentice with a power of $48\%$.}

{In scenarios G, I-2 and in some of the censoring rates in I-3, the Peto--Peto and Pepe--Fleming tests tend to be with the highest power. In contrast, in many of the scenarios in which the survival functions cross, their power are much lower than our tests. For example, in Scenario K-1, $n=400$, and $25\%$ censoring rate, the power of KONP is  $89\%$, while Pepe--Fleming power is $42\%$ and Peto--Peto is $5\%$.}

To conclude, for the $2$-sample setting, in most of the non-proportional hazards settings, our proposed KONP tests tend to be more powerful than the other tests, and the differences between $S_P$ and $S_{LR}$ are very small, if any.

Results of settings with $K>2$ can be found in Table S1 and Figure S1 of the Supplementary Material. Based on these results we conclude that all the tests are valid, as the empirical size of the tests are reasonably close to $0.05$. For $K=3$ and scenarios D and J-2, we see similar results to those shown with $K=2$, as KONP tests are often much more powerful than the logrank and Peto--Peto test. For example, for scenario D with $n=200$ and $25\%$ censoring rates in all the three groups, the powers of KONP tests are approximately $92\%$, while of Peto--Peto is $49\%$, and of logrank is $18\%$.

\begin{figure}[!h]
	\caption{Empirical power the 2-samples settings under proportional hazards or close to proportionality: {Scenario L is with proportional hazards. M is close to proportionality but with substantial differences at early times. N--Q are close to proportionality and model (\ref{hr_yp}).}   }
	\centering
	\label{prop_res}
		\includegraphics[width=1.1\textwidth]{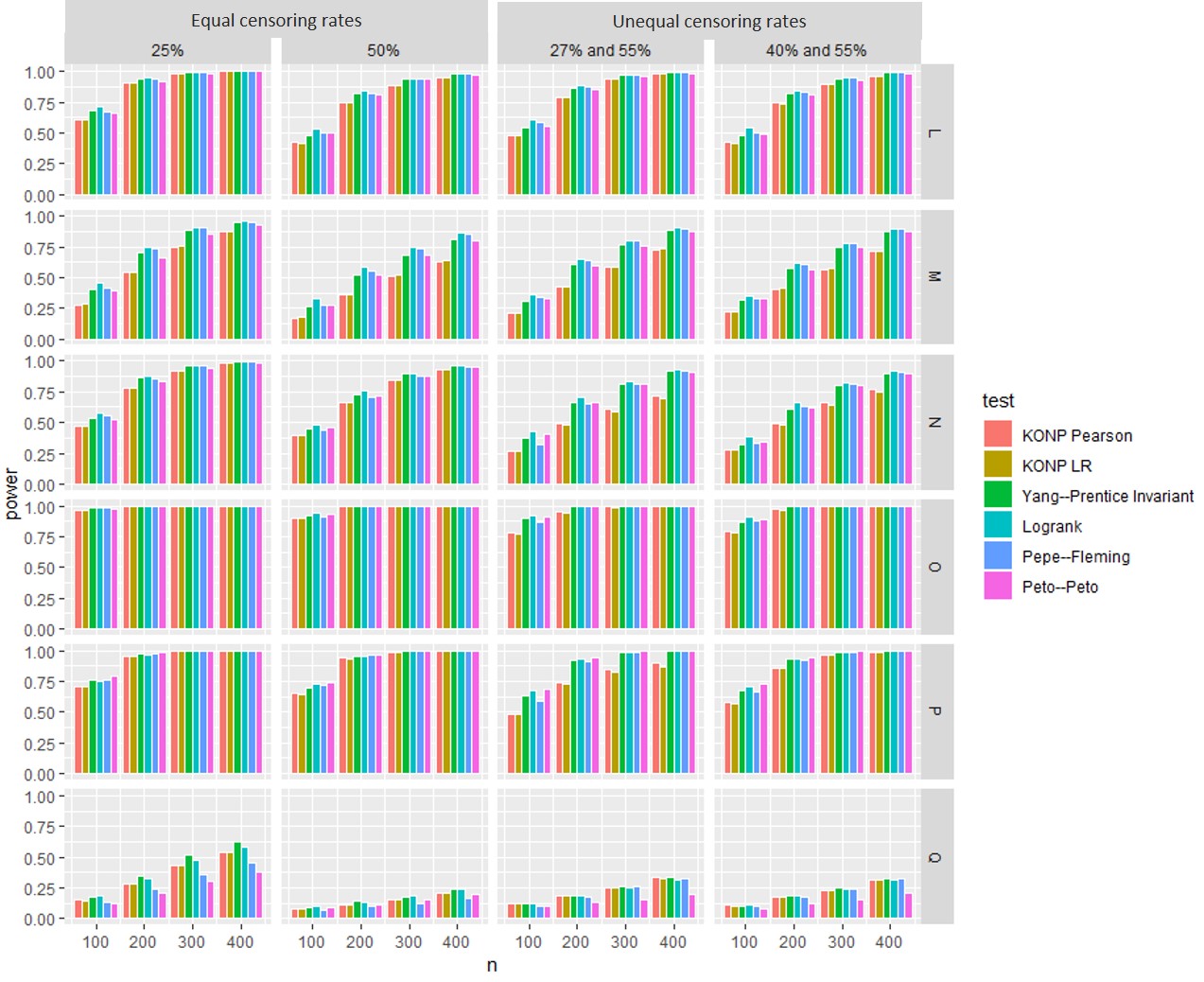}
\end{figure}

{Figure \ref{prop_res} summarizes the power of the 2-sample tests under proportional hazards or close to proportionality. Under these settings the logrank test is often with the highest power among the invariant tests, as expected. The invariant version of Yang--Prentice test is similar to the logrank test, and the proposed KONP tests, sometimes, have less power.}

\subsection{A Robust Approach}
Figures \ref{ref_scen_fig} and \ref{our_scen_fig} of the main text and Table S2 of the Supplementary Material indicate that under the non-proportional hazards scenarios and among the invariant tests, usually the proposed KONP tests are with the largest power. Under proportional hazards or close to proportionality, usually the logrank test is with the largest power among the invariant tests. The invariant version of Yang--Prentice test is similar to the logrank test under proportional hazards settings, since their model contains the proportional hazards model ($\theta_1=\theta_2$).    

In case one is interested in a robust powerful test under non-proportional or proportional hazards, the principle of minimum p-value could be adopted based on the elegant Cauchy-combination test of \citet{liu2018cauchy}, which is similar to the test based on the minimum $p$-value. Denote the $p$-values of our KONP tests by $p\mbox{-value}_P$ and $p\mbox{-value}_{LR}$, respectively, and the $p$-value of the logrank test by $p\mbox{-value}_{lgrnk}$. Then, based on \citet{liu2018cauchy} we define a new test statistic
\begin{equation}
Cau = [\tan \{ (0.5- p\mbox{-value}_P) \pi \}]/3 + [\tan \{ (0.5-p\mbox{-value}_{LR}) \pi \}]/3  + [\tan \{ (0.5-p\mbox{-value}_{lgrnk}) \pi \}]/3\, , \nonumber
\end{equation} 
and its $p$-value is
\begin{equation}
p\mbox{-value}_{Cau} = 0.5 - (\arctan Cau)/\pi \, . \nonumber
\end{equation}
The candidate tests to be included in ${Cau}$ are powerful tests under non-proportional (i.e.  KONP tests) or proportional hazards (i.e. logrank test and the invariant version of Yang--Prentice test). Due to the similarity in power performances between the logrank test and the invariant version of Yang--Prentice test, under proportional or close to proportional hazards, and due to the high computational burden of Yang--Prentice invariant test, only the logrank test is included.

{Figure S3 of the Supplementary Materials provides the empirical Type-I error of the two KONP tests, the logrank test and the robust $Cau$ test. Evidently, the size of the tests are reasonably close to 0.05. Figures S4-S5 of the Supplementary Materials summarize the empirical power, based on 1000 replications, of the two KONP tests, the logrank test and the test statistic $Cau$. Often, the $Cau$ test loses some power comparing to the largest power among KONP and logrank, but the loss is relatively small. Table S4 of the Supplementary Materials provides the power values of these tests under all the studied alternatives, along with two additional combined test: the test of Lee~\cite{lee2007} which is based on the maximum of two weighted logrank test statistics; and the MaxCombo test~\cite{lin2019} which is based on the maximum of the logrank and three weighted logrank test statistics. (See Supplementary Materials for details.) Lee and MaxCombo tests
	perform very similarly.
	 In general, our $Cau$ test outperforms Lee and the MaxCombo tests, in terms of power, in all the sub-scenarios (i.e., various sample size and censoring patterns) of type B, C, H, J-1, and J-2; and in some of the sub-scenarios of A, F, I-1, I-2, I-3, K-1, K-2, K-3, and P. All of these are non-proportional hazards scenarios. Under proportional hazards or close to proportionality, i.e., settings L, M, N O, P and Q, Lee's test or MaxCombo outperform $Cau$. Interestingly, the cases where Lee or MaxCombo tests are with higher power than $Cau$, the power loss by using $Cau$ is relatively small; while this is not always the case when $Cau$ outperforms Lee or MaxCombo. For example, under Setting B with $n=200$, the power of $Cau$ equals 0.919 while the power values of Lee and MaxCombo are 0.499 and 0.450, respectively.} 

Our R package KONPsurv \citep{schl2019} applies the above robust test $Cau$ as well.

\section{Real data examples}
\subsection{The Gastrointestinal Tumor Data}
\label{read_data_section}
\begin{figure}[]
	\caption{Gastrointestinal Tumor Study: KM Curves}
	\centering\label{gastric_plot}
	\includegraphics[scale=0.5]{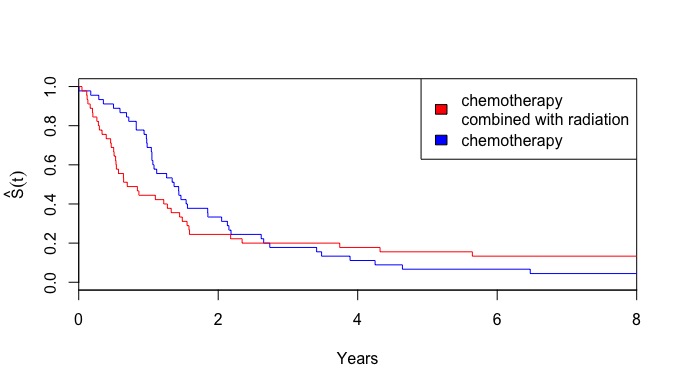}
\end{figure}

The Gastrointestinal Tumor Study Group \citep{schein1982comparison}
compared chemotherapy with combined chemotherapy and radiation therapy, in the treatment of locally unresectable gastric cancer. This dataset was used in \citet{yang2010improved} to demonstrate the utility of their test. Each treatment arm had 45 patients, and two observations of the chemotherapy group and six of the combination group were censored. The primary outcome measure was time to death. The KM survival curves of time to death, in each treatment group are provided in Figure~\ref{gastric_plot}. To apply the Yang--Prentice test, we considered chemotherapy as the control group and chemotherapy plus radiation therapy as the treatment group, which is named Yang--Prentice 1. The Yang--Prentice test with reversed group labeling is denoted by Yang--Prentice 2. Table \ref{gastric_pv} shows the $p$-values of testing for equality of the survival curves of time to death, of the two treatment groups, against a two-sided alternative, with each of the tests considered in the simulation study. For our tests and the Yang-Prentice invariant test, $10$ imputations and $10^4$ permutations for each imputation, were used. Evidently, the smallest $p$-values are observed under our proposed KONP tests and the Cauchy-combination test $Cau$.

\begin{figure}[]
	\caption{Urothelial Carcinoma Data: KM Curves}
	\centering\label{Uro_plot}
	\includegraphics[scale=0.6]{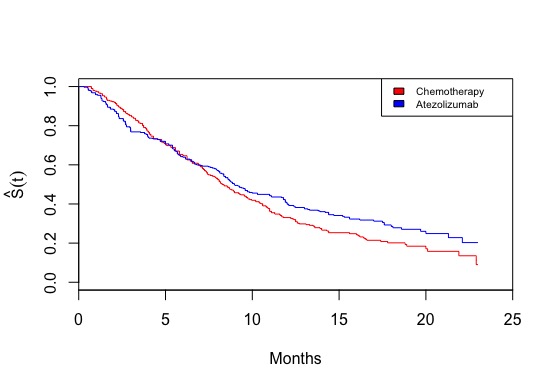}
\end{figure}

{\subsection{Urothelial Carcinoma}
Few options exist for patients with locally advanced or metastatic urothelial carcinoma after progression with platinum-based chemotherapy. Powel et al. \cite{powles2018atezolizumab}  aimed to assess the safety and efficacy of atezolizumab versus chemotherapy in this patient population.
Their study consists of a  multi-center, open-label, phase 3
randomised controlled trial conducted at 217 academic medical centers and community oncology practices mainly in Europe, North America, and the Asia-Pacific region. The primary endpoint was overall survival. Figure S3 of the Supplementary Material of \citet{powles2018atezolizumab} provides the overall survival KM curves of atezolizumab versus chemotherapy based on 316 and 309 patients, respectively. Although the detailed survival data are unavailable, inspired by \cite{Roych2019} we used Guyot et al. \cite{Guyot2012} algorithm that maps from digitised curves back to KM data, by finding numerical
solutions to the inverted KM equations, using available information on number of events and numbers at risk. The DigitizeIt software was used for reading the coordinates of the KM curves from the published graph. Figure \ref{Uro_plot} provides the KM curves by treatment arm, and the last column of Table \ref{gastric_pv} shows the $p$-values of testing for equality of the survival curves of time to death against a two-sided alternative. Evidently, the smallest $p$-values are observed under our proposed KONP tests and the second best is the test of Lee. Our Cauchy combined test also performs robustly.
}

\section{Discussion and Conclusions}
\label{Discussion}
The proposed KONP tests are based on partition of the sample-space into two subsets, corresponding to three intervals, as of the HHG test \citep{heller2013consistent}. An extensive simulation study shows that when the hazard curves are non-proportional, the KONP tests are often more powerful than all the other tests. In particular, the proposed test is even more powerful than the Yang--Prentice test, under their model with non-proportional hazards. The simulation results show very little differences in power, if any, between the Pearson chi-squared test statistic and the log-likelihood ratio statistic. Since the chi-squared statistic is slightly more powerful, this test statistic is recommended. 

Other partitions and summary statistics can be easily adopted. In particular, one may consider the extended Anderson-Darling tests of \citet{thas2004extension} and \citet{heller2016consistent} with higher sample-space partitions and test statistics that aggregate over all partitions by summation or maximization. Nevertheless, for non-censored data, \citet{heller2016consistent} showed by simulations (see their Table 1), that increasing the number of partitions can improve power over the HHG 2-sample test under settings in which the density functions intersect 4 times or more. Otherwise, the HHG 2-sample test tends to be more powerful. Figure \ref{density_fig} in Appendix \ref{long_table} displays  the densities of the 17 non-proportional hazards scenarios studied in this work. Evidently, the survival scenarios considered by others and by us are of less than 4 intersections. Simulation study of the Anderson-Darling test statistic with sample-space partition of two intervals, yields lower power than the proposed KONP tests. A comprehensive comparison with other sample-space partitions and aggregations could be a topic of future research. 

This work suggests tests that accommodate right-censored data. Since the tests are based on the KM estimator, it seems that a modification to left truncation might be possible. However, additional work is  required to modify the imputation-permutation approach for left-truncated data. 

Implementation of our tests, KONP-P, KONP-LR and $Cau$, is available in the R package KONPsurv \citep{schl2019}, which can be freely downloaded from CRAN.

\section*{Acknowledgement}
The authors gratefully acknowledge support from the NIH (R01CA189532) and the U.S.-Israel Binational
Science Foundation (2016126), in carrying out this work.

\section*{Supplementary materials}
\label{SM}
Supplementary materials include: (1) The proof of the theorem. (2) Description of some of the tests included in the simulation study. (3) Description and results of $K$-sample simulation settings with $K=3,4,5$.  (4) Plots of the empirical power results of the robust test $Cau$. 
(5) The exact empirical power used for generating all the plots in the main text and the Supplementary Material file.

\clearpage

\begin{table}
	\centering
	\caption{Computation Time (seconds) }
	\begin{tabular}{cccccc}
		Censoring rates  & \multicolumn{5}{c}{$n$} \\
		of groups 1 and 2 & 100 & 200 & 300 & 400 & 1000 \\
		(25\%,25\%) & 1.7 & 7.1 & 16.5 & 30.0 & 204.2 \\
		(50\%,50\%) & 0.9 & 3.5 & 8.0 & 14.5 & 97.3 \\
		(27\%,55\%) & 0.9 & 3.8 & 8.8 & 16.1 & 114.1 \\
		(40\%,55\%) & 0.8 & 3.2 & 7.5 & 13.7 & 93.8 \\
	\end{tabular}\label{running_time_table}
\end{table}

\begin{table}
	\centering
	\caption{Unequal Censoring Distributions}
	\begin{tabular}{cccc}
		Group      & Small difference            & Substantial difference     \\ 
		1 & $ \min \{U(a,b) , Exp(\theta_1) \}$ & $ \min \{U(a,b) , Exp(\theta_1) \}$  \\ 
		2 & $ \min \{U(a,b) , Exp(\theta_2) \}$ & $ U(a,b)$   \\ 
	\end{tabular}\label{unequal_cen}
\end{table}

\begin{table}
	\centering
	\caption{Examples: Gastrointestinal Tumor Study (GST) and Urothelial Carcinoma Data (UCD)}
	\begin{tabular}{lcc}
		Test                    &  GST $p$-value & UCD $p$-value\\ 
		KONP Pearson        & 0.0109 & 0.0049 \\ 
		KONP LR             & 0.0108 & 0.0049 \\
		Cauchy combination $Cau$             & 0.0164 & 0.0071 \\
		Yang--Prentice 1        & 0.0304 & 0.0186 \\ 
		Yang--Prentice 2        & 0.0800 & 0.0252 \\ 
		Yang--Prentice Invariant & 0.0479 & 0.0219 \\ 
		Logrank                 & 0.6350  & 0.0673 \\ 
		Pepe--Fleming            & 0.9464 & 0.2362 \\ 
		Peto--Peto               & 0.0465 &0.3807 \\
		Lee  	 & 0.0968	& 0.0054\\
		MaxCombo 	 & 0.0908	&  0.0061\\ 
	\end{tabular}\label{gastric_pv}
\end{table}

\begin{thebibliography}{18}
	\providecommand{\natexlab}[1]{#1}
	\providecommand{\url}[1]{\texttt{#1}}
	\expandafter\ifx\csname urlstyle\endcsname\relax
	\providecommand{\doi}[1]{doi: #1}\else
	\providecommand{\doi}{doi: \begingroup \urlstyle{rm}\Url}\fi
	
	\bibitem[Yang and Prentice(2010)]{yang2010improved}
	Yang S and Prentice RL.
	\newblock Improved logrank-type tests for survival data using adaptive weights.
	\newblock \emph{Biometrics}, 66\penalty0 (1):\penalty0 30--38, 2010.
	
	\bibitem[Peto and Peto(1972)]{peto1972asymptotically}
	Peto R and Peto J.
	\newblock Asymptotically efficient rank invariant test procedures.
	\newblock \emph{Journal of the Royal Statistical Society. Series A}, pages
	185--207, 1972.
	
	\bibitem[Pepe and Fleming(1989)]{pepe1989weighted}
	Pepe MS and Fleming TR.
	\newblock Weighted {K}aplan--{M}eier statistics: a class of distance tests for
	censored survival data.
	\newblock \emph{Biometrics}, pages 497--507, 1989.
	
	\bibitem[Pepe and Fleming(1991)]{pepe1991weighted}
	Pepe MS and Fleming TR.
	\newblock Weighted {K}aplan--{M}eier statistics: {L}arge sample and optimality
	considerations.
	\newblock \emph{Journal of the Royal Statistical Society. Series B}, pages
	341--352, 1991.
	
	\bibitem[Yang and Prentice(2005)]{yang2005semiparametric}
	Yang S and Prentice RL.
	\newblock Semiparametric analysis of short-term and long-term hazard ratios
	with two-sample survival data.
	\newblock \emph{Biometrika}, 92\penalty0 (1):\penalty0 1--17, 2005.
	
	\bibitem[Darling(1957)]{darling1957kolmogorov}
	Darling DA.
	\newblock The {K}olmogorov--{S}mirnov, {C}ramer--von {M}ises tests.
	\newblock \emph{Annals of Mathematical Statistics}, 28\penalty0 (4):\penalty0
	823--838, 1957.
	
	\bibitem[Pettitt(1976)]{pettitt1976two}
	Pettitt AN.
	\newblock A two-sample {A}nderson--{D}arling rank statistic.
	\newblock \emph{Biometrika}, 63\penalty0 (1):\penalty0 161--168, 1976.
	
	\bibitem[Scholz and Stephens(1987)]{scholz1987k}
	Scholz FW and Stephens MA.
	\newblock $k$-sample {A}nderson--{D}arling tests.
	\newblock \emph{Journal of the American Statistical Association}, 82\penalty0
	(399):\penalty0 918--924, 1987.
	
	\bibitem[Thas and Ottoy(2004)]{thas2004extension}
	Thas O and Ottoy JP.
	\newblock An extension of the {A}nderson--{D}arling $k$-sample test to
	arbitrary sample space partition sizes.
	\newblock \emph{Journal of Statistical Computation and Simulation}, 74\penalty0
	(9):\penalty0 651--665, 2004.
	
	\bibitem[Heller et~al.(2013)Heller, Heller, and Gorfine]{heller2013consistent}
	Heller R, Heller Y, and Gorfine M.
	\newblock A consistent multivariate test of association based on ranks of
	distances.
	\newblock \emph{Biometrika}, 100\penalty0 (2):\penalty0 503--510, 2013.
	
	\bibitem[Heller et~al.(2016)Heller, Heller, Kaufman, Brill, and
	Gorfine]{heller2016consistent}
	Heller R, Heller Y, Kaufman S, Brill B, and Gorfine M.
	\newblock Consistent distribution-free ${K}$-sample and independence tests for
	univariate random variables.
	\newblock \emph{Journal of Machine Learning Research}, 17\penalty0
	(1):\penalty0 978--1031, 2016.
	
	\bibitem[Wang et~al.(2010)Wang, Lagakos, and Gray]{wang2010testing}
	Wang R, Lagakos SW, and Gray RJ.
	\newblock Testing and interval estimation for two-sample survival comparisons
	with small sample sizes and unequal censoring.
	\newblock \emph{Biostatistics}, 11\penalty0 (4):\penalty0 676--692, 2010.
	
	\bibitem[Uno et~al.(2015)Uno, Tian, Claggett, and Wei]{uno2015versatile}
	Uno H, Tian L, Claggett B, and Wei LJ.
	\newblock A versatile test for equality of two survival functions based on
	weighted differences of {K}aplan--{M}eier curves.
	\newblock \emph{Statistics in medicine}, 34\penalty0 (28):\penalty0 3680--3695,
	2015.
	
	\bibitem[Yang(2019)]{yang2019interim}
	Yang S.
	\newblock Interim monitoring using the adaptively weighted log-rank test in
	clinical trials for survival outcomes.
	\newblock \emph{Statistics in medicine}, 38\penalty0 (4):\penalty0 601--612,
	2019.
	
	\bibitem[Lin et~al.(1993)Lin, Wei, and Ying]{lin1993checking}
	Lin DY, Wei LJ, and  Ying Z.
	\newblock Checking the cox model with cumulative sums of martingale-based
	residuals.
	\newblock \emph{Biometrika}, 80\penalty0 (3):\penalty0 557--572, 1993.
		
	\bibitem[Liu and Xie(2019)]{liu2018cauchy}
     Liu Y and  Xie J.
	\newblock Cauchy combination test: a powerful test with analytic p-value
	calculation under arbitrary dependency structures.
	\newblock \emph{Journal of the American Statistical Association} 1-18, 2019.
	
	\bibitem[Lee (2007)]{lee2007}
	Seung-Hwan L.
	\newblock On the versatility of the combination of the weighted log-rank statistics.
	\newblock \emph{Computational Statistics \& Data Analysis} ,
	51(12), 6557--6564, 2007.
	
	\bibitem[Lin et~al.(2019)]{lin2019}
	Lin RS, et al. 
	\newblock Alternative Analysis Methods for Time to Event Endpoints under Non-proportional Hazards: A Comparative Analysis.
	\newblock \emph{arXiv preprint} arXiv:1909.09467, 2019.
	
	
	\bibitem[Schlesinger and Gorfine(2019)]{schl2019}
	Schlesinger M and Gorfine M.
	\newblock Konp tests: Powerful k-sample tests for right-censored data.
	\newblock 2019.
	\newblock R package version 1.0.1.
	
	\bibitem[Schein and Group(1982)]{schein1982comparison}
	Schein  PS and Gastrointestinal Tumor~Study Group.
	\newblock A comparison of combination chemotherapy and combined modality
	therapy for locally advanced gastric carcinoma.
	\newblock \emph{Cancer}, 49\penalty0 (9):\penalty0 1771--1777, 1982.
	
	\bibitem[Powel et~al.(2018)]{powles2018atezolizumab}
	Powles T et al.
	\newblock Atezolizumab versus chemotherapy in patients with platinum-treated locally advanced or metastatic urothelial carcinoma (IMvigor211): a multicentre, open-label, phase 3 randomised controlled trial.
	\newblock \emph{The Lancet} 391: 748--757, 2018.
	
	\bibitem[Roychoudhury et~al.(2019)]{Roych2019}
	Roychoudhury S et al.
	\newblock Robust Design and Analysis of Clinical Trials With Non-proportional Hazards: A Straw Man Guidance from a Cross-pharma Working Group.
	\newblock \emph{arXiv preprint} arXiv:1908.07112, 2019.
	
	\bibitem[Guyot et~al.(2012)]{Guyot2012}
	Guyot P, Ades AE, Ouwens MJ, and Welton NJ.
	\newblock Enhanced secondary analysis of survival data: reconstructing the data from published Kaplan-Meier survival curves.
	\newblock \emph{BMC medical research methodology}, (12):\penalty1 9, 2012.
	
\end{thebibliography}

\setcounter{table}{0}
\setcounter{figure}{0}
\renewcommand{\thetable}{A\arabic{table}}
\renewcommand{\thefigure}{A\arabic{figure}}
\newpage
\section{Appendix}
\subsection{Detailed Description of the Simulation Settings}
\label{long_table}
{\footnotesize
\begin{longtable}{|l|l|l|l|l|}
\caption{Description of the Non-Proportional Hazards Simulation Scenarios\label{scenario_description}}
\\
\hline
\multicolumn{1}{|c|}{Scenario} & \multicolumn{1}{c|}{\begin{tabular}[c]{@{}c@{}}Failure time distribution \\ and reference\end{tabular}}     & \multicolumn{1}{c|}{\begin{tabular}[c]{@{}c@{}}Graphical \\ description\end{tabular}}    
 & \multicolumn{2}{c|}{Censoring distribution}                                                                                                                                                                                         \\ \hline
Null                           & \begin{tabular}[c]{@{}l@{}}$F_1(t) = log$-$Logistic(1,1)$ \\ \\ $F_2(t) = log$-$Logistic(1,1)$\\ \\  Yang and Prentice (2010)\end{tabular} & \begin{minipage}{.20\textwidth} \ \includegraphics[width=\linewidth, height=25mm]{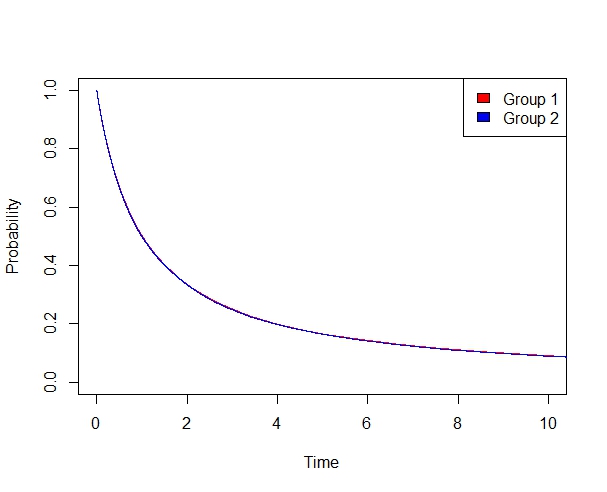} \end{minipage} & \multicolumn{2}{l|}{\begin{tabular}[c]{@{}l@{}}equal:\\ $C \sim log$-$Normal(\alpha,0.5)$\\ $\alpha=(1.1,0)$ \\ \\ unequal: $(a,b)=(0,10)$\\ $(\theta_1,\theta_2)=(0.85,0.25)$\end{tabular}} \\ \hline

A                           & \begin{tabular}[c]{@{}l@{}}$
F_1(t) = \left\{
        \begin{array}{ll}
            Weibull(0.849,10) & \quad t \le 3 \\
            U(3,50.625) & \quad 3<t \le 33 \\
            Weibull(0.849,10) & \quad t>33
        \end{array} 
    \right.$ \\ \\ $F_2 = Weibull(0.849,10)$\\ \\ Difference in middle times, \\ Uno et al. (2015)\end{tabular} & \begin{minipage}{.20\textwidth} \ \includegraphics[width=\linewidth, height=25mm]{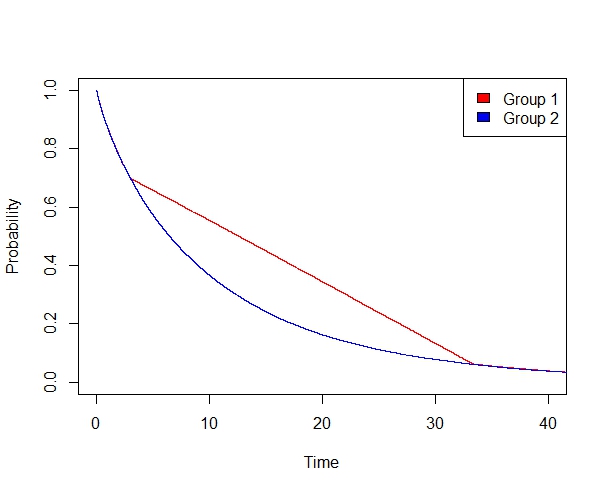} \end{minipage} & \multicolumn{2}{l|}{\begin{tabular}[c]{@{}l@{}}equal: $C \sim Weibull(\alpha,\beta)$\\ $(\alpha,\beta)=(18,16), (1.5,9)$ \\ \\ unequal: $(a,b)=(2,30)$\\ $(\theta_1,\theta_2)=(0.06,0.04)$\end{tabular}} \\ \hline

B                           & \begin{tabular}[c]{@{}l@{}}$
F_1(t) = \left\{
        \begin{array}{ll}
            U(0,50) & \quad t \le 3 \\
            U(3,12.347) & \quad 3<t \le 8 \\
            Weibull(0.849,10) & \quad t>8
        \end{array} 
    \right.$ \\ \\ $F_2(t) = Weibull(0.849,10)$\\ \\ Difference in early times, \\ Uno et al. (2015)\end{tabular} & \begin{minipage}{.20\textwidth} \ \includegraphics[width=\linewidth, height=25mm]{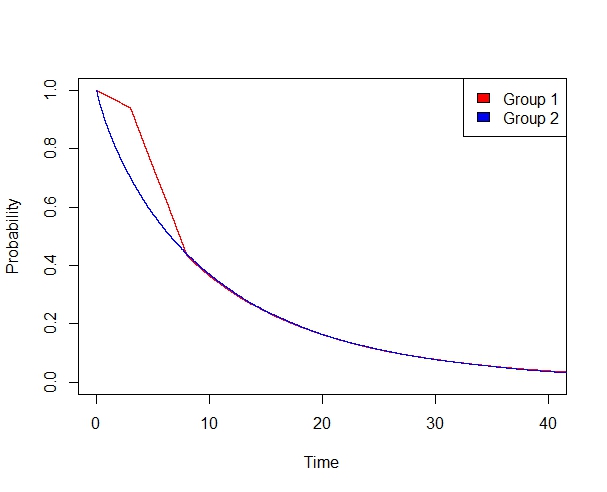} \end{minipage} & \multicolumn{2}{l|}{\begin{tabular}[c]{@{}l@{}}equal: $C \sim Weibull(\alpha,\beta)$\\ $(\alpha,\beta)=(10,15), (3,7.5)$ \\ \\ unequal: $(a,b)=(2,30)$\\ $(\theta_1,\theta_2)=(0.06,0.04)$\end{tabular}} \\ \hline

C                           & \begin{tabular}[c]{@{}l@{}}$
F_1(t) = \left\{
        \begin{array}{ll}
            Exp(0.5) & \quad t \le 0.57 \\
            Exp(1.5) & \quad 0.57<t \le 1.1 \\
            Exp(1) & \quad t>1.1
        \end{array} 
    \right.$ \\ $
F_2(t) = \left\{
        \begin{array}{ll}
            Exp(1.5) & \quad t \le 0.56 \\
            Exp(2/9) & \quad 0.56<t \le 1.1 \\
            Exp(1) & \quad t>1.1
        \end{array} 
    \right.$\\  Difference in early times, \\ Pepe and Fleming (1989; 1991)\end{tabular} & \begin{minipage}{.20\textwidth} \ \includegraphics[width=\linewidth, height=25mm]{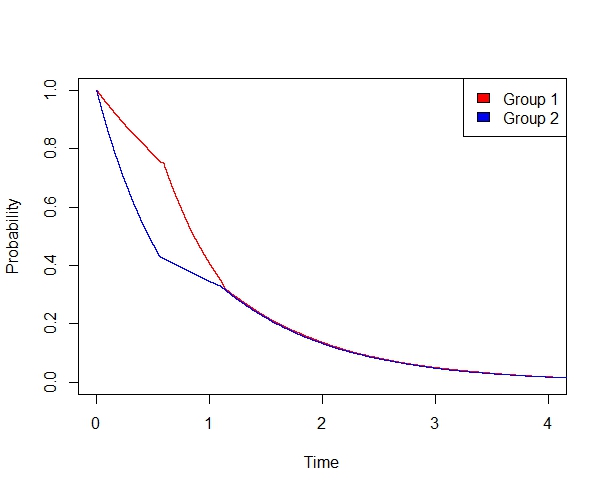} \end{minipage} & \multicolumn{2}{l|}{\begin{tabular}[c]{@{}l@{}}equal: $C \sim U(\alpha,\beta)$\\ $(\alpha,\beta)=(1,2), (0,1.6)$ \\ \\ unequal: $(a,b)=(0.01,3)$\\ $(\theta_1,\theta_2)=(0.5,0.8)$\end{tabular}} \\ \hline    

D                           & \begin{tabular}[c]{@{}l@{}}$
F_1(t) = \left\{
        \begin{array}{ll}
            Exp(0.5) & \quad t \le 0.44 \\
            Exp(0.1) & \quad 0.44<t \le 1.05 \\
            Exp(1.5) & \quad 1.05<t \le 1.47 \\
            Exp(1) & \quad t>1.47
        \end{array} 
    \right.$ \\  $
F_2(t) = \left\{
        \begin{array}{ll}
            Exp(1.5) & \quad t \le 0.38 \\
            Exp(0.1) & \quad 0.38<t \le 1.02 \\
            Exp(0.5) & \quad 1.02<t \le 1.47 \\
            Exp(1) & \quad t>1.47
        \end{array} 
    \right.$\\  Difference in early times, \\ Pepe and Fleming (1989; 1991)\end{tabular} & \begin{minipage}{.20\textwidth} \ \includegraphics[width=\linewidth, height=25mm]{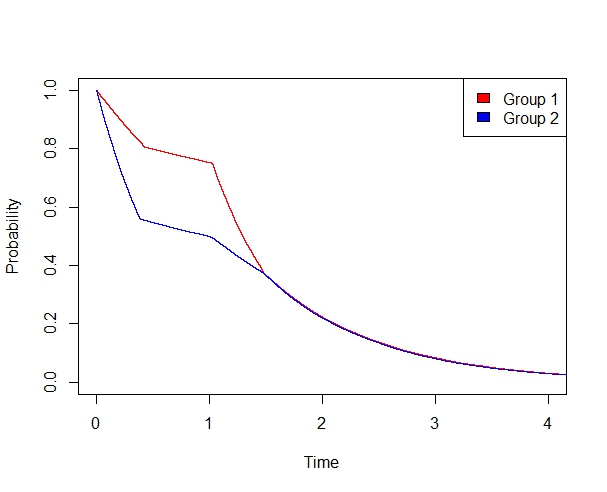} \end{minipage} & \multicolumn{2}{l|}{\begin{tabular}[c]{@{}l@{}}equal: $C \sim U(\alpha,\beta)$\\ $(\alpha,\beta)=(1.1,3), (0.1,2.1)$ \\ \\ unequal: $(a,b)=(0.5,3.5)$\\ $(\theta_1,\theta_2)=(0.5,0.3)$\end{tabular}} \\ \hline    
    
E                           & \begin{tabular}[c]{@{}l@{}}$ F_1(t) = Exp(1)$ \\ \\ $
F_2(t)= \left\{
        \begin{array}{ll}
            Exp(1) & \quad t \le 0.3 \\
            Exp(2) & \quad t>0.3 
        \end{array} 
    \right.$\\ \\Proportional difference in late times, \\ Pepe and Fleming (1989; 1991)\end{tabular} & \begin{minipage}{.20\textwidth} \ \includegraphics[width=\linewidth, height=25mm]{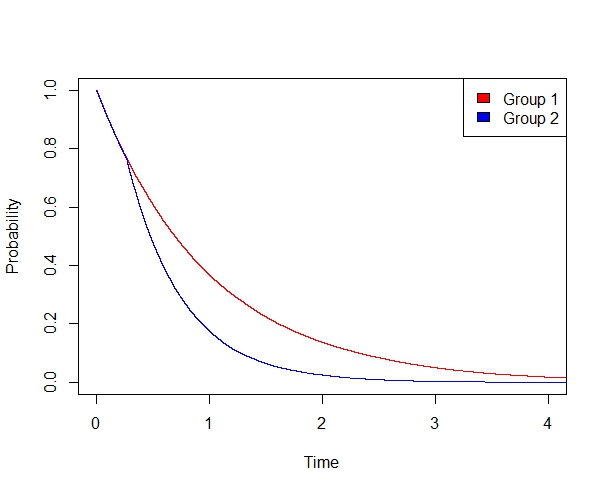} \end{minipage} & \multicolumn{2}{l|}{\begin{tabular}[c]{@{}l@{}}equal: $C \sim U(\alpha,\beta)$\\ $(\alpha,\beta)=(0.9,1.2), (0.1,1.1)$ \\ \\ unequal: $(a,b)=(0.01,2.3)$\\ $(\theta_1,\theta_2)=(0.5,0.8)$\end{tabular}} \\ \hline

F                           & \begin{tabular}[c]{@{}l@{}}$
F_1(t)=1-\{1+\exp(2)t\}^{-\exp(1)}$  \\ \\
$F_2(t)= log$-$Logistic(1,1)$ \\ \\ Yang and Prentice Model (1), \\ Yang and Prentice (2010)\end{tabular} & \begin{minipage}{.20\textwidth} \ \includegraphics[width=\linewidth, height=25mm]{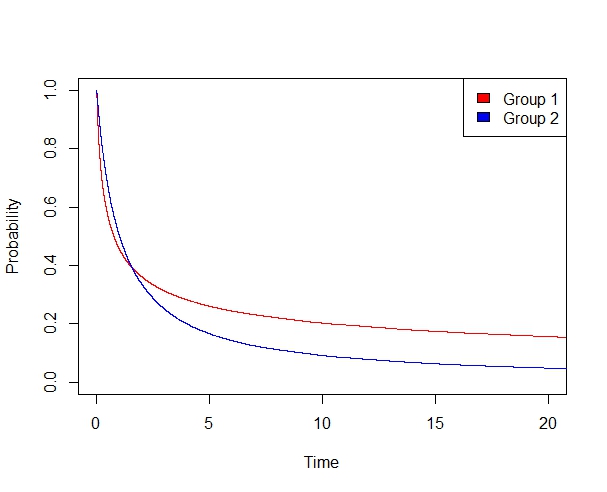} \end{minipage} & \multicolumn{2}{l|}{\begin{tabular}[c]{@{}l@{}}equal:\\ $C \sim log$-$Normal(\alpha,0.5)$\\ $\alpha=(1.35,0.01)$ \\ \\ unequal: $(a,b)=(0.5,5)$\\ $(\theta_1,\theta_2)=(0.7,0.25)$\end{tabular}} \\ \hline    

G                           & \begin{tabular}[c]{@{}l@{}}$
F_1(t)=1-\{1+t/ \exp(2)\}^{-\exp(1)}$  \\ \\
$F_2(t) = log$-$Logistic(1,1)$ \\ \\ Yang and Prentice Model (1), \\ Yang and Prentice (2010)\end{tabular} & \begin{minipage}{.20\textwidth} \ \includegraphics[width=\linewidth, height=25mm]{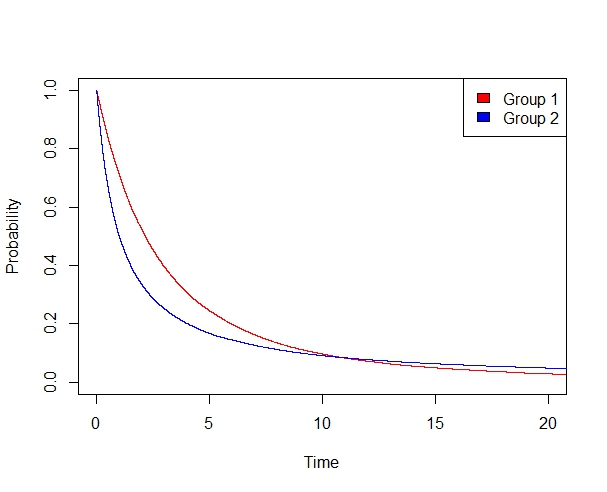} \end{minipage} & \multicolumn{2}{l|}{\begin{tabular}[c]{@{}l@{}}equal:\\ $C \sim log$-$Normal(\alpha,0.5)$\\ $\alpha=(1.4,0.4)$ \\ \\ unequal: $(a,b)=(0.5,7)$\\ $(\theta_1,\theta_2)=(0.2,0.4)$\end{tabular}} \\ \hline  

H                           & \begin{tabular}[c]{@{}l@{}}$
F_1(t)= Exp (1)$  \\ \\
$F_2(t)= \left\{
        \begin{array}{ll}
            Exp(2) & \quad t \le 0.5 \\
            Exp(0.5) & \quad 0.5<t \le 1.5 \\
            Exp(2) & \quad t>1.5
        \end{array} 
    \right.$ \\ \\  U shape hazard ratio, \\ Yang and Prentice (2010)\end{tabular} & \begin{minipage}{.20\textwidth} \ \includegraphics[width=\linewidth, height=25mm]{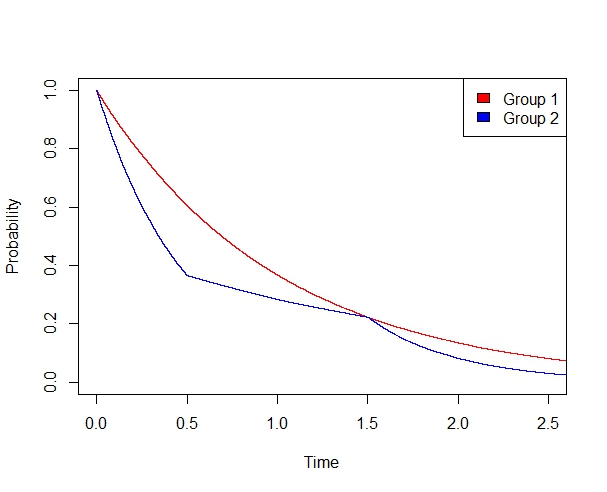} \end{minipage} & \multicolumn{2}{l|}{\begin{tabular}[c]{@{}l@{}}equal:\\ $C \sim log$-$Normal(\alpha,0.5)$\\ $\alpha=(0.25,-0.7)$ \\ \\ unequal: $(a,b)=(0,2.5)$\\ $(\theta_1,\theta_2)=(0.5,0.25)$\end{tabular}} \\ \hline  
    
I-1                          & \begin{tabular}[c]{@{}l@{}}$
F_1(t)= Exp (1)$  \\ \\
$F_2(t)=1-\{4\exp(t)+3\}^{-0.5}$ \\ \\ Yang and Prentice Model (1), \\ \end{tabular} & \begin{minipage}{.20\textwidth} \ \includegraphics[width=\linewidth, height=23mm]{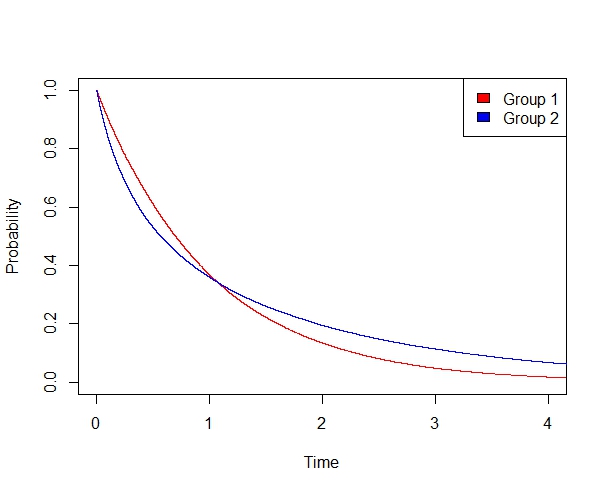} \end{minipage} & \multicolumn{2}{l|}{\begin{tabular}[c]{@{}l@{}}equal: $C \sim Exp(\lambda)$\\ $\lambda=(0.32,1)$ \\ \\ unequal: $(a,b)=(0,4)$\\ $(\theta_1,\theta_2)=(0.8,0.4)$\end{tabular}} \\ \hline  

I-2                          & \begin{tabular}[c]{@{}l@{}}$
F_1(t)= Exp (1)$  \\ \\
$F_2(t)=1-\big[\{\exp(t)+3\}/4\big]^{-2}$ \\ \\ Yang and Prentice Model (1) \end{tabular} & \begin{minipage}{.20\textwidth} \ \includegraphics[width=\linewidth, height=23mm]{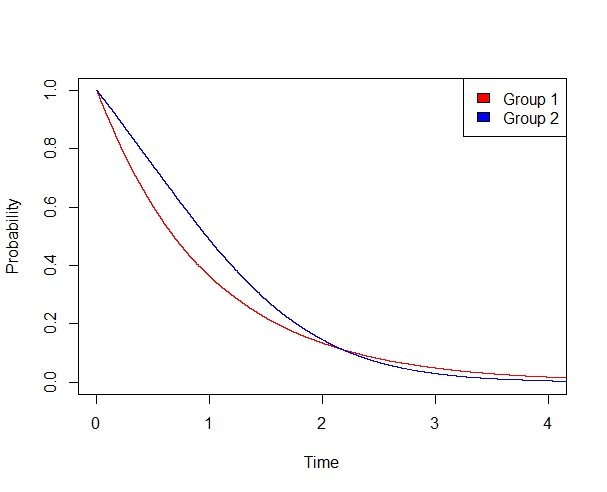} \end{minipage} & \multicolumn{2}{l|}{\begin{tabular}[c]{@{}l@{}}equal: $C \sim Exp(\lambda)$\\ $\lambda=(0.3,0.85)$ \\ \\ unequal: $(a,b)=(0,4)$\\ $(\theta_1,\theta_2)=(0.8,0.25)$\end{tabular}} \\ \hline  

I-3                           & \begin{tabular}[c]{@{}l@{}}$
F_1(t)=\{5/(0.5t+5)\}^5$  \\ \\
$F_2(t)=log$-$Logistic(1,1)$ \\  \\ Yang and Prentice Model (1) \end{tabular} & 
\begin{minipage}{.20\textwidth} \ \includegraphics[width=\linewidth, height=25mm]{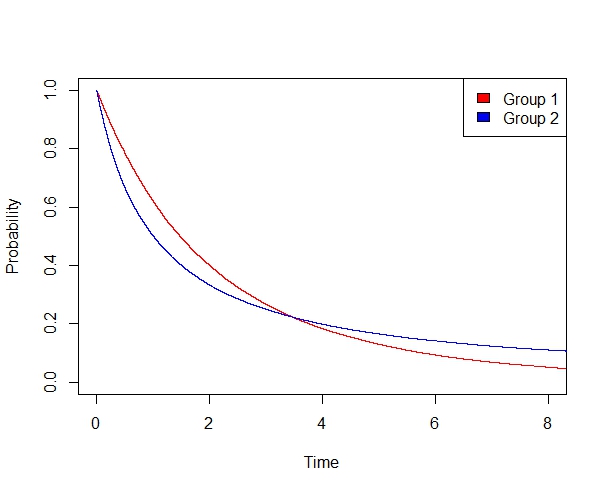} \end{minipage} & \multicolumn{2}{l|}{\begin{tabular}[c]{@{}l@{}}equal:\\ $C \sim log$-$Normal(\alpha,0.5)$\\ $\alpha=(1.2,0.25)$ \\ \\ unequal: $(a,b)=(0,10)$\\ $(\theta_1,\theta_2,)=(0.4,0.25)$\end{tabular}} \\ \hline

J-1                           & \begin{tabular}[c]{@{}l@{}}$
F_1(t)= Exp (1)$  \\ \\
$F_2(t)=  \left\{
        \begin{array}{ll}
            Exp(2) & \quad t \le 0.25 \\
            Exp(0.6) & \quad t>0.25
        \end{array} 
    \right.$ \\  \\ strong monotone hazards ratio\\ assumption does not hold \end{tabular} & \begin{minipage}{.20\textwidth} \ \includegraphics[width=\linewidth, height=25mm]{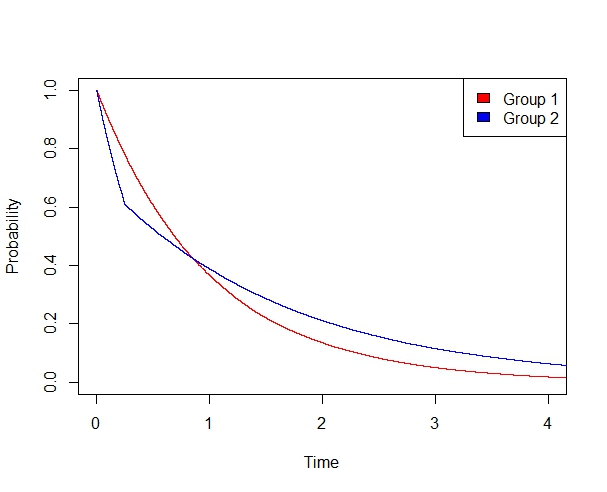} \end{minipage} & \multicolumn{2}{l|}{\begin{tabular}[c]{@{}l@{}}equal: $C \sim Exp(\lambda)$\\ $\lambda=(0.3,1)$ \\ \\ unequal: $(a,b)=(0,4)$\\ $(\theta_1,\theta_2)=(0.9,0.5)$\end{tabular}} \\ \hline  

J-2                           & \begin{tabular}[c]{@{}l@{}}$
F_1(t)= Exp (1)$  \\ \\
$F_2(t)= \left\{
        \begin{array}{ll}
            Exp(1) & \quad t \le 0.1 \\
            Exp(1.7) & \quad 0.1<t \le 0.45 \\
            Exp(0.5) & \quad t>0.45
        \end{array} 
    \right.$ \\  \\ strong monotone hazards ratio\\ assumption does not hold \end{tabular} & \begin{minipage}{.20\textwidth} \ \includegraphics[width=\linewidth, height=25mm]{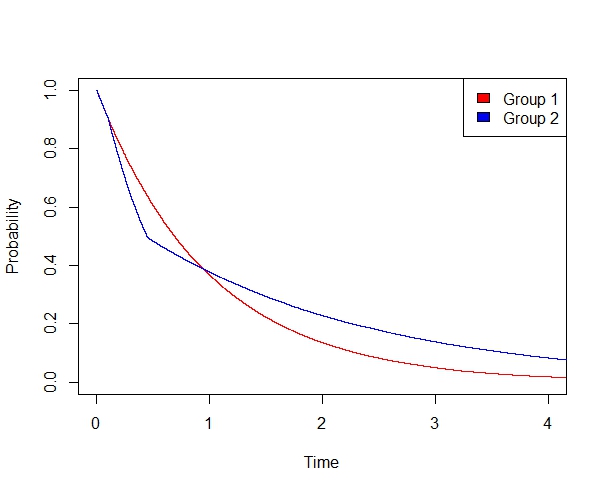} \end{minipage} & \multicolumn{2}{l|}{\begin{tabular}[c]{@{}l@{}}equal: $C \sim Exp(\lambda)$\\ $\lambda=(0.3,1)$ \\ \\ unequal: $(a,b)=(0,4)$\\ $(\theta_1,\theta_2)=(0.9,0.5)$\end{tabular}} \\ \hline  
    
J-3                           & \begin{tabular}[c]{@{}l@{}}$
F_1(t)= Exp (1)$   \\ \\
$F_2(t)= \left\{
        \begin{array}{ll}
            \exp\{0.5t^2-t\} & \quad t \le 1 \\
            \exp\{-0.5t^2+t-1\} & \quad t>1
        \end{array} 
    \right.$ \\  \\ strong monotone hazards ratio\\ assumption does not hold \end{tabular} & \begin{minipage}{.20\textwidth} \ \includegraphics[width=\linewidth, height=25mm]{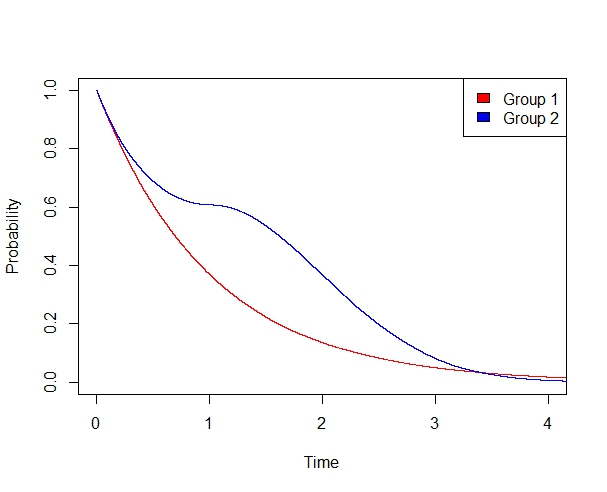} \end{minipage} & \multicolumn{2}{l|}{\begin{tabular}[c]{@{}l@{}}equal: $C \sim Exp(\lambda)$\\ $\lambda=(0.25,0.75)$ \\ \\ unequal: $(a,b)=(0,5)$\\ $(\theta_1,\theta_2)=(0.9,0.3)$\end{tabular}} \\ \hline  

K-1                           & \begin{tabular}[c]{@{}l@{}}$
F_1(t)= Exp (1)$  \\ \\
$F_2(t)=\{4\exp(1.7t)-3\}^{-1/3.4}$ \\  \\ strong monotone hazards ratio \end{tabular} & \begin{minipage}{.20\textwidth} \ \includegraphics[width=\linewidth, height=25mm]{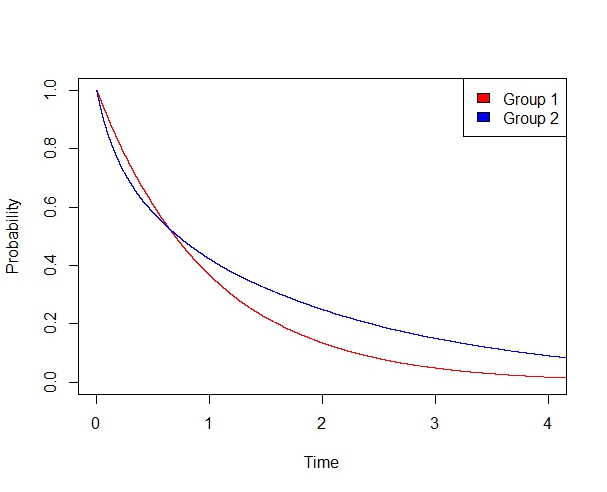} \end{minipage} & \multicolumn{2}{l|}{\begin{tabular}[c]{@{}l@{}}equal: $C \sim Exp(\lambda)$\\ $\lambda=(0.3,0.9)$ \\ \\ unequal: $(a,b)=(0,5)$\\ $(\theta_1,\theta_2)=(0.9,0.3)$\end{tabular}} \\ \hline  

K-2                           & \begin{tabular}[c]{@{}l@{}}$
F_1(t)=Exp (1)$  \\ \\
$F_2(t)=\{8\exp(2t)-7\}^{-0.25}$ \\  \\ strong monotone hazards ratio \end{tabular} & \begin{minipage}{.20\textwidth} \ \includegraphics[width=\linewidth, height=25mm]{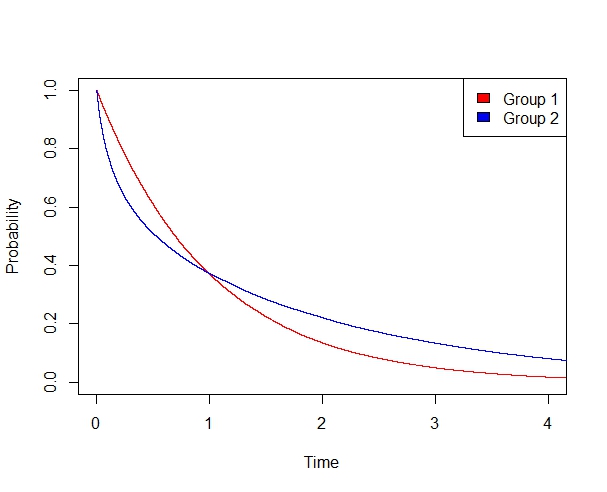} \end{minipage} & \multicolumn{2}{l|}{\begin{tabular}[c]{@{}l@{}}equal: $C \sim Exp(\lambda)$\\ $\lambda=(0.3,1)$ \\ \\ unequal: $(a,b)=(0,5)$\\ $(\theta_1,\theta_2)=(0.9,0.45)$\end{tabular}} \\ \hline  

K-3                           & \begin{tabular}[c]{@{}l@{}}$
F_1(t)= Exp (0.4)$  \\ \\
$F_2(t)=\big[\{2\exp(t)+64\}/66\big]^{-2.5}$ \\  \\ strong monotone hazards ratio \end{tabular} & \begin{minipage}{.20\textwidth} \ \includegraphics[width=\linewidth, height=25mm]{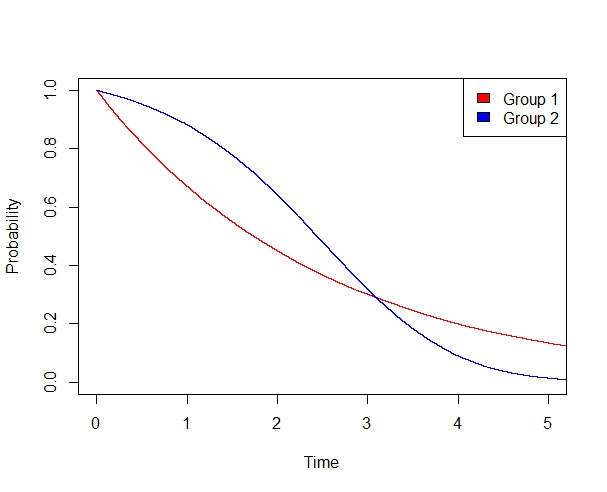} \end{minipage} & \multicolumn{2}{l|}{\begin{tabular}[c]{@{}l@{}}equal: $C \sim Exp(\lambda)$\\ $\lambda=(0.13,35)$ \\ \\ unequal: $(a,b)=(0,10)$\\ $(\theta_1,\theta_2)=(0.3,0.15)$\end{tabular}} \\ \hline  
 \end{longtable}}

\newpage
{\footnotesize
	\begin{longtable}[c]{|l|l|l|l|l|}
		\caption{Additional 2-samples settings: proportional hazards and close to proportional hazards 
			\label{prop_design}}\\
		\hline
		\multicolumn{1}{|c|}{Scenario} & \multicolumn{1}{c|}{\begin{tabular}[c]{@{}c@{}}Failure time distribution \\ and reference\end{tabular}}     & \multicolumn{1}{c|}{Graphical description}                                                              & \multicolumn{2}{c|}{Censroing distribution}                                                                                                                                                                                         \\ \hline
		L                           & \begin{tabular}[c]{@{}l@{}}$F_1(t) = Weibull(0.849,20)$ \\ \\ $F_2(t) = Weibull(0.849,10)$\\  \\ Proportional hazards, \\ Uno et al. (2015)\end{tabular} & 
		\begin{minipage}{.20\textwidth} \ \includegraphics[width=\linewidth, height=25mm]{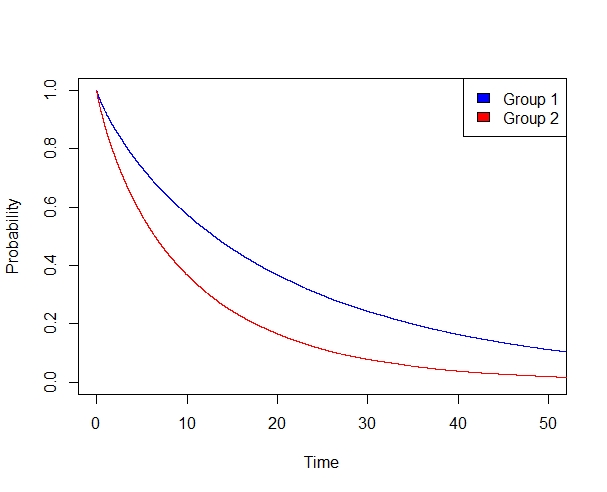} \end{minipage} & \multicolumn{2}{l|}{\begin{tabular}[c]{@{}l@{}}equal:\\ $C \sim Weibull(\alpha,\beta)$\\ $(\alpha,\beta)=(5,24), (1.5,12)$ \\ \\ unequal: $(a,b)=(0,40)$\\ $(\theta_1,\theta_2)=(0.025,0.05)$\end{tabular}} \\ \hline
		
		M                           & \begin{tabular}[c]{@{}l@{}}$
			F_1(t) = \left\{
			\begin{array}{ll}
			Weibull(4,1) & \quad t \le 0.5 \\
			Weibull(2,1.5) & \quad t>0.5
			\end{array} 
			\right.$ \\ \\ $F_2 = \left\{
			\begin{array}{ll}
			Weibull(2.2,1) & \quad t \le 0.5 \\
			Weibull(1.5,1.5) & \quad t>0.5
			\end{array} 
			\right.$\\  \\ Substantial difference in early times, \\ Pepe and Fleming (1989; 1991)\end{tabular} & \begin{minipage}{.20\textwidth} \ \includegraphics[width=\linewidth, height=24mm]{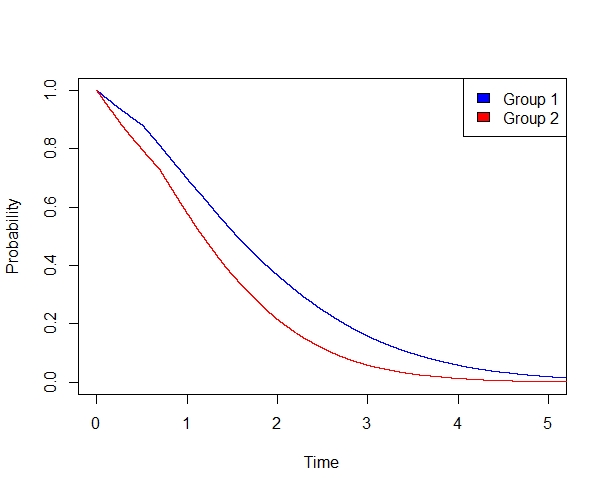} \end{minipage} & \multicolumn{2}{l|}{\begin{tabular}[c]{@{}l@{}}equal: $C \sim Weibull(\alpha,\beta)$\\ $(\alpha,\beta)=(0.9,5.5), (0.35,3.4)$ \\ \\ unequal: $(a,b)=(0,4.5)$\\ $(\theta_1,\theta_2)=(0.25,0.14)$\end{tabular}} \\ \hline
		
		N                           & \begin{tabular}[c]{@{}l@{}}$
			F_1(t)=1-(1+t)^{exp(-0.5)}$  \\ \\
			$F_2(t) = log$-$Logistic(1,1)$ \\  \\ Yang and Prentice model, \\ Yang and Prentice (2010)\end{tabular} & \begin{minipage}{.20\textwidth} \ \includegraphics[width=\linewidth, height=24mm]{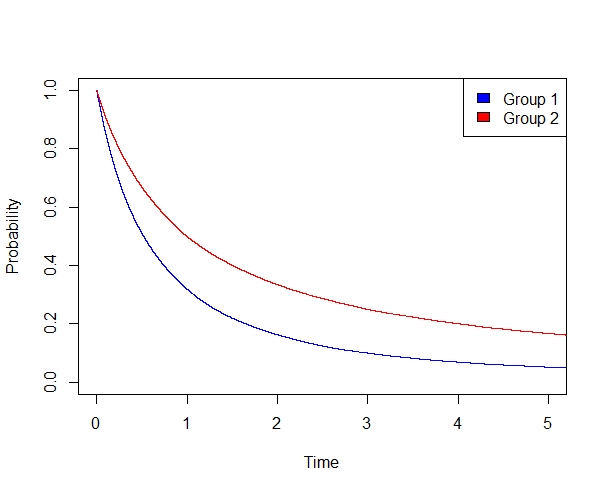} \end{minipage} & \multicolumn{2}{l|}{\begin{tabular}[c]{@{}l@{}}equal:\\ $C \sim log$-$Normal(\alpha,0.5)$\\ $\alpha=(0.75,-0.1)$ \\ \\ unequal: $(a,b)=(0,12)$\\ $(\theta_1,\theta_2)=(1.5,0.4)$\end{tabular}} \\ \hline  
		
		O                           & \begin{tabular}[c]{@{}l@{}}$
			F_1(t)=1-(1+t)^{exp(-1)}$  \\ \\
			$F_2(t) = log$-$Logistic(1,1)$ \\  \\ Yang and Prentice model, \\ Yang and Prentice (2010)\end{tabular} & \begin{minipage}{.20\textwidth} \ \includegraphics[width=\linewidth, height=24mm]{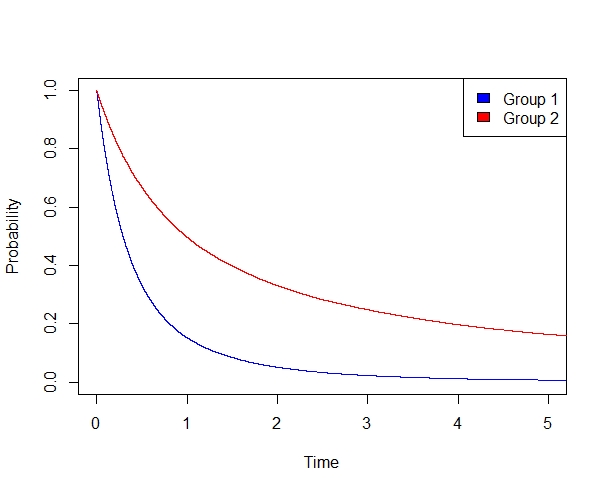} \end{minipage} & \multicolumn{2}{l|}{\begin{tabular}[c]{@{}l@{}}equal:\\ $C \sim log$-$Normal(\alpha,0.5)$\\ $\alpha=(-0.6,-1.8)$ \\ \\ unequal: $(a,b)=(0,8)$\\ $(\theta_1,\theta_2)=(2,0.5)$\end{tabular}} \\ \hline  
		
		P                           & \begin{tabular}[c]{@{}l@{}}$
			F_1(t)=1-\{1+t\exp(1)\}^{-1}$  \\ \\
			$F_2(t) = log$-$Logistic(1,1)$ \\  \\ Yang and Prentice model, \\ Yang and Prentice (2010)\end{tabular} & \begin{minipage}{.20\textwidth} \ \includegraphics[width=\linewidth, height=25mm]{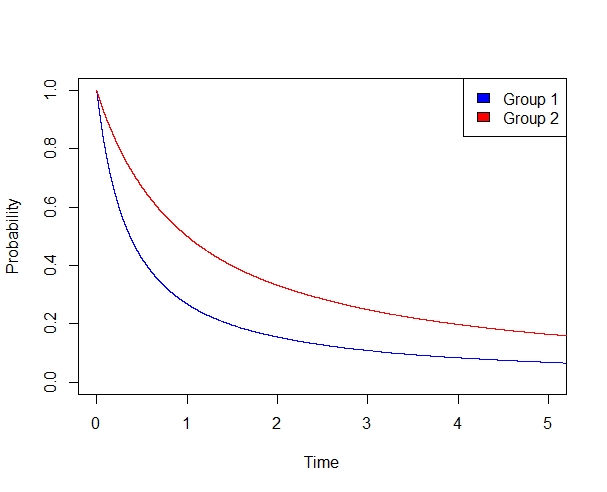} \end{minipage} & \multicolumn{2}{l|}{\begin{tabular}[c]{@{}l@{}}equal:\\ $C \sim log$-$Normal(\alpha,0.5)$\\ $\alpha=(0.6,-0.5)$ \\ \\ unequal: $(a,b)=(0,8)$\\ $(\theta_1,\theta_2)=(2,0.5)$\end{tabular}} \\ \hline  
		
		Q                           & \begin{tabular}[c]{@{}l@{}}$
			F_1(t)=1-\{1+t/ \exp(1)\}^{-exp(1)}$  \\ \\
			$F_2(t) = log$-$Logistic(1,1)$ \\  \\ Yang and Prentice model, \\ Yang and Prentice (2010)\end{tabular} & \begin{minipage}{.20\textwidth} \ \includegraphics[width=\linewidth, height=25mm]{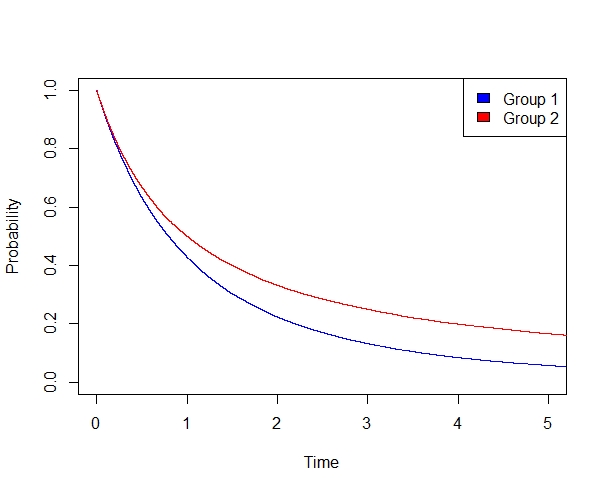} \end{minipage} & \multicolumn{2}{l|}{\begin{tabular}[c]{@{}l@{}}equal:\\ $C \sim log$-$Normal(\alpha,0.5)$\\ $\alpha=(0.7,-0.15)$ \\ \\ unequal: $(a,b)=(0,8)$\\ $(\theta_1,\theta_2)=(0.9,0.3)$\end{tabular}} \\ \hline  
	\end{longtable}
}

\begin{figure}[!h]
\caption{Density plots of the 17 simulation scenarios: red (green) - density function of treatment group 1 (2).}
\centering
\label{density_fig}
\includegraphics[width=0.65\textwidth]{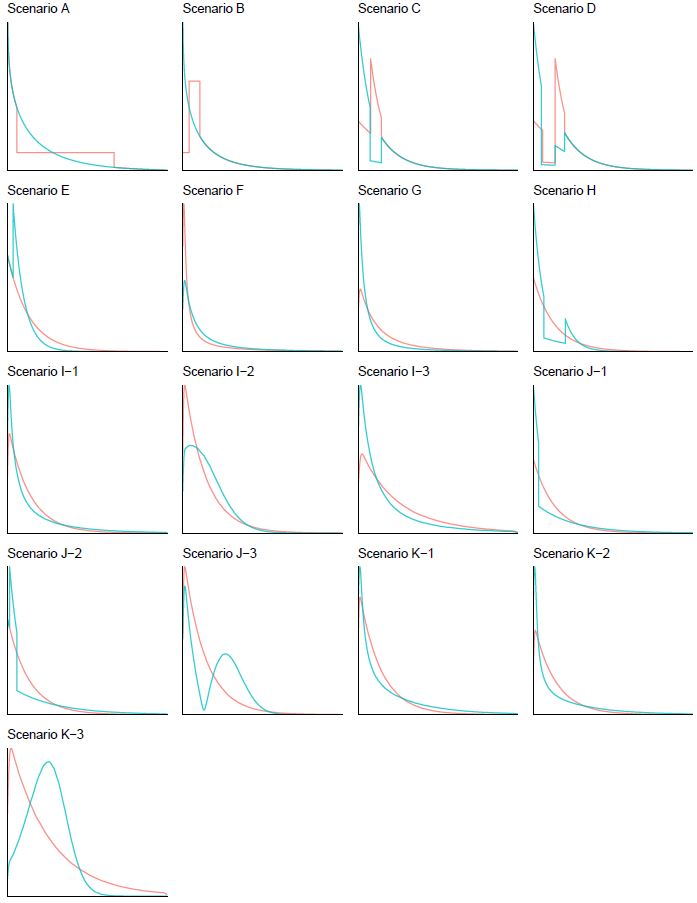}
\end{figure}

\renewcommand{\thetable}{S\arabic{table}}
\renewcommand{\thefigure}{S\arabic{figure}}

\newpage
\section{Supplementary Materials}

This Supplementary Material file includes, the proof of the theorem, additional simulation results and additional details of the simulation results summarized in the paper. 

\begin{center}
	\textbf{Proof of the theorem for a continuous survival time}
\end{center}
\label{proof_contin}
For simplicity, we show the proof using the Pearson chi-squared test statistic. The proof using the likelihood ratio test statistic is very similar and therefore omitted. The reasoning is based on  the proofs of Heller et al. (2013). We show in the following that for an arbitrary fixed $\alpha \in (0,1)$, if $H_0$ is false then $\lim_{n\to\infty} \pr \big(Q-q_{1-\alpha} >0 \big)=1$, where $q_{1-\alpha}$ is the $1-\alpha$ quantile of the test statistic under the null distribution.

Assume $X\in \mathbb{R}^+$ has a continuous distribution given $Y$, denoted by $f_{X|Y}(\cdot|\cdot)$, and let $f_X^*(x)=\sum_{k=1}^K \pi_k f_{X|Y}(x|k)$, which is not necessarily the true marginal distribution. If $H_0$ is false, there exists at least one pair of points $(x_0,g)$ such that without loss of generality $f_{X|Y}(x_0|g)>f_X^*(x_0)$. Assume (for the moment) that $\pr(C>x_0|Y=k)>0$ for all $k=1,\ldots,K$ and let $d(x,x_0)=|x-x_0|$. Since $f_{X|Y}(\cdot |g)$ and $f_X^*(\cdot)$ are continuous, there exist a radius, $R>0$, and a set 
$$
\mathcal{B}=\{x: d(x,x_0)<R \}
$$ 
with positive probability, such that for $x \in \mathcal{B}$, $f_{X|Y}(x | g)>f_X^*(x)$ and $\pr(C>x+3R|Y=k)>0$ for all $k=1,\ldots,K$. The last condition guarantees that with positive probability $S_P(i,j)$ with $n_{ij}=n$ (namely, the table consists of all the groups) is observed, where $Y_i=g$, and $X_i, X_j \in \mathcal{B}$. Moreover,
$$\min_{\mathcal{B}} \{f_{X|Y}(x | g) - f_X^*(x)\} > 0 \, ; $$ 
denote this minimum by $c$. Put
\begin{eqnarray*}
	\mathcal{B}_1 &= & \{ x : d(x, x_0)< R/8 \}\ ,\\ 
	\mathcal{B}_2 &=& \{ x : 3R/8 < d(x, x_0) < R/2 \}\,  ,
\end{eqnarray*}
and let $p_1=\pr (X \in \mathcal{B}_1,Y=g)$, $p_2 = \pr(X \in \mathcal{B}_2)$,
$p_3 = \pr(C>x_0+R,Y=g)$, and $p_4 = \min_{k=1,\ldots,K}\pr(C>x_0+R,Y=k)$. Therefore, we expect to have at least $n^2 p_1 p_2 p_3 p_4$ pairs $(i,j)$ such that $\{X_i\in \mathcal{B}_1,Y_i=g, \Delta_i=1\}$ and $\{X_j\in \mathcal{B}_2, \Delta_j=1\}$. Consider such a pair.

Stute and Wang (1993) showed that the KM estimator converges almost surely to the true survival function under random censorship. Therefore, uniformly almost surely,
\begin{eqnarray*}
	&&\lim_{n\to\infty} \left\{ \frac{A_{11}^*(i,j)}{n-2} - \pi_g \int_{\mathcal{B}_3}f_{X|Y}(x|Y=g)dx \right\} = 0 \, ,\\
	&&\lim_{n\to\infty} \left\{ \frac{A_{1\cdot}^*(i,j)}{n-2} - \pi_g\ \right\} = 0 \, ,\\
	&&\lim_{n\to\infty} \left\{ \frac{A_{\cdot1}^*(i,j)}{n-2} - \int_{\mathcal{B}_3}f_X^*(x)dx \right\}= 0 \, ,
\end{eqnarray*}
where 
$$\mathcal{B}_3=\{x: d(x,X_i)<d(X_i,X_j) \, .\}
$$ 
Since 
$$S_P(i,j)=\sum_{m=1,2}\sum_{l=1,2}\frac{\{A_{ml}^*(i,j)-A_{m\cdot}^*(i,j)A_{\cdot l}^*(i,j)/(n-2)\}^2}{A_{m\cdot}^*(i,j)A_{\cdot l}^*(i,j)/(n-2)} \ ,$$
it is enough to look at the term with $l=m=1$ in $S_P(i,j)$, that is,
$$S_{P_1}(i,j)=\frac{\{A_{11}^*(i,j)-A_{1\cdot}^*(i,j)A_{\cdot 1}^*(i,j)/(n-2)\}^2}{A_{1\cdot}^*(i,j)A_{\cdot 1}^*(i,j)/(n-2)} \ .$$
It follows that $S_P(i,j)\geq S_{P_1}(i,j)$ and hence that our test statistic satisfies 
$$ 
Q\geq\frac{1}{N}\sum_{k=1}^{K}\sum_{T_i\in \mathcal{A}_k}\sum_{j=1,j\neq i}^{n} S_{P_1}(i,j) \Delta_i \Delta_j I(T_j\le \tau_k) I(2T_i-T_j\le \tau_k) \equiv Q_1\, .
$$
By Slutzky's theorem and the continuous mapping theorem, in probability 
\begin{eqnarray*}
	\lim_{n\to\infty} \frac{S_{P_1}(i,j)}{n-2} &=& \lim_{n\to\infty} \frac{1}{n-2} \frac{\{A_{11}^*(i,j)-A_{1\cdot}^*(i,j)A_{\cdot 1}^*(i,j)/(n-2)\}^2}{A_{1\cdot}^*(i,j)A_{\cdot 1}^*(i,j)/(n-2)}   \\ 
	&=& \frac{\pi_g\big[\int_{\mathcal{B}_3}\{f_{X|Y}(x|Y=g) - f_X^*(x)\}dx \big]^2}{\int_{\mathcal{B}_3}f_X^*(x)dx} \, ,
\end{eqnarray*}
and we show that this limit can be bounded from below by a positive constant that does not depend on $(i,j)$. Indeed, it can be shown that $\mathcal{B}_3 \subseteq \mathcal{B}$ and $\mathcal{B}_1 \subseteq \mathcal{B}_3$, by the triangle inequality (see Heller et al. for details). Therefore, 
$$
\pi_g\Big[\int_{\mathcal{B}_3}\{f_{X|Y}(x|Y=g) - f_X^*(x)\}dx \Big]^2 \geq \pi_g \Big\{c \int_{\mathcal{B}_3}dx\Big\}^2 \geq \pi_g \Big\{c \int_{\mathcal{B}_1}dx\Big\}^2 \equiv c'
$$
since $\int_{\mathcal{B}_3}f_X^*(x)dx \leq 1$, it follows that $S_{P_1}(i,j)/(n-2)$ converges in probability to a positive constant greater than $c'>0$. Therefore, $S_{P_1}^*(i,j)>(n-2)c'$ with probability going to $1$ as $n\to\infty$. As a result, $Q_1>N^{-1} n^2(n-2)c'p_1p_2p_3p_4$ with probability going to $1$ as $n\to\infty$. Therefore,
$$
\lim_{n\to\infty}\pr \big\{Q - N^{-1}n^2(n-2)c'p_1p_2p_3p_4 > 0 \big\}=1 \ .
$$
Since $N<n^2$, there exist a constant $\lambda>0$ such that
$\lim_{n\to\infty}\pr \big(Q - \lambda n > 0 \big) =1$.
Under the null hypothesis, for a large enough sample size $n$, $S_P(i, j)$ follows the $\chi^2$ distribution with one degree of freedom. Therefore, the null expectation of $Q$, which is an average of $N$ different $S_P$'s, is approximately 1 and the null variance is bounded above by 2. Consequently, $\lim_{n\to\infty}(\lambda n-q_{1-\alpha})>0$ and  $\lim_{n\to\infty} \pr \big(Q - q_{1-\alpha} > 0 \big)=1$.
Finally, for simplicity of presentation, we required that $\pr(C>x_0|Y=k)>0$ for all $k=1,\ldots,K$. However this is only required  for the two different groups, $g$ and $m$. The proof for the discrete failure time variable $X$ is done in much the same way.\\

\begin{center}
	\textbf{Description of the tests included in the simulation study}
\end{center}
Let 
$$
G^{\rho,\gamma} = \sqrt{\frac{n_1+n_2}{n_1n_2}} \int_0^{\infty} 
\left\{ \widehat{S}(t-) \right\}^{\rho} \left\{ 1- \widehat{S}(t-) \right\}^{\gamma} 
\frac{\overline{Y}_1(t)\overline{Y}_2(t)}{\overline{Y}_1(t)+\overline{Y}_2(t)}
\left\{ \frac{d \overline{N}_1(t)}{\overline{Y}_1(t)} - \frac{d \overline{N}_2(t)}{\overline{Y}_2(t)}\right\}
$$
where $\overline{N}_j(t)$ is the number of failures before or at time $t$ in group $j$, $\overline{Y}_j(t)$  is the number at risk at time $t$ in group $j$, $j=1,2$, and  $\widehat{S}$ is the Kaplan-Meier estimator based on the pooled data. Also let
\begin{eqnarray}
\widehat{\sigma}_{lm} &=& {\frac{n_1+n_2}{n_1n_2}} \int_0^{\infty} 
\left\{ \widehat{S}(t-) \right\}^{\rho_l} \left\{ 1- \widehat{S}(t-) \right\}^{\gamma_l}
\left\{ \widehat{S}(t-) \right\}^{\rho_m} \left\{ 1- \widehat{S}(t-) \right\}^{\gamma_m} \nonumber \\
&&\frac{\overline{Y}_1(t)\overline{Y}_2(t)}{\overline{Y}_1(t)+\overline{Y}_2(t)}
\left\{ 1 - \frac{\Delta \overline{N}_1(t) + \Delta \overline{N}_2(t) -1}{\overline{Y}_1(t)+\overline{Y}_2(t)}  \right\}
\frac{d\{\overline{N}_1(t) +  \overline{N}_2(t)\}}{\overline{Y}_1(t)+\overline{Y}_2(t)}\nonumber 
\end{eqnarray}
where $\Delta \overline{N}_j(t) = \overline{N}_j(t) - \overline{N}_j(t-)$, $j=1,2$, 
$(\rho_1,\gamma_1)=(0,0)$ (the logrank test statistic), $(\rho_2,\gamma_2)=(1,0)$, 
$(\rho_3,\gamma_3)=(0,1)$, and $(\rho_4,\gamma_4)=(1,1)$. Then, four standardized statistics are defined by
$$
Z_k = G^{\rho_k,\gamma_k} / \sqrt{\widehat{\sigma}_{kk}} \;\;\; , k=1,\ldots,4 \,.
$$ 
The test statistic of Lee (2007) is defined by $$\max\{|Z_2|,|Z_3|\}$$ and the MaxCombo test statistic is  $$\max\{|Z_1|,|Z_2|,|Z_3|,|Z_4|\}.$$ The $p$values can  be easily calculated based on the asymptotic multivariate normal distribution of $$(G^{\rho_1,\gamma_1},G^{\rho_2,\gamma_2},G^{\rho_3,\gamma_3},G^{\rho_4,\gamma_4})$$
under the null hypothesis, i.e. with a mean zero and a covariance matrix that can be consistently estimated by $\widehat{\sigma}_{lm}$. The R package mvtnorm was used. 

The Pepe-Fleming weighted KM test (Pepe and Fleming, 1989) uses the following weight function
$$
\frac{n \widehat{G}_1(t)\widehat{G}_2(t)}{n_1 \widehat{G}_1(t) + n_2 \widehat{G}_1(t)}
$$ 
where $\widehat{G}_j$ is the KM estimator of the time to censoring in group $j$, $j=1,2$. Details od the variance estimator can be found in Pepe and Fleming (1989). Peto-Peto weighted KM test (Peto and Peto, 1972) uses a weight function that is very close to the pooled KM estimator.

\footnotesize
\clearpage
{\footnotesize
	\begin{longtable}[c]{|l|l|l|l|}
		\caption{$K$-sample scenarios with $K=3,4,5$: under the null hypothesis and based on scenarios D and J-2 from the main text
			\label{K_sample_design}}\\
		\hline
		\multicolumn{1}{|c|}{Scenario} &\multicolumn{1}{c|}{Failure time}                                                              & \multicolumn{2}{c|}{Censroing distribution}                                                                                                                                                                                         \\ \hline
		
		\begin{tabular}[c]{@{}l@{}} Null \\ $K=3$ \end{tabular}                      &  \begin{tabular}[c]{@{}l@{}}$F_1(t) = F_2(t) = F_3(t) =$\\ $log$-$Logistic(1,1)$ \end{tabular} & \multicolumn{2}{l|}{\begin{tabular}[c]{@{}l@{}}equal:\\ $C \sim log$-$Normal(\alpha,0.5)$\\ $\alpha=(1.1,0)$ \\ \\ 40\% and 55\%:\\ $C_1,C_2\sim\min \{Exp(0.85),U(0,10)\}$ \\ $C_3 \sim U(0,10)$ \\ \\ 27\% and 55\%: \\ $C_1 \sim \min \{Exp(0.85),U(0,10)\}$\\
				$C_2 \sim \min \{Exp(0.25),U(0,10)\}$ \\ $C_3\sim U(0,10)$ \end{tabular}} \\ \hline
		
		\begin{tabular}[c]{@{}l@{}} Null \\ $K=4$ \end{tabular}                                     &  \begin{tabular}[c]{@{}l@{}}$F_1(t) = F_2(t) = F_3(t) = F_4(t) = $ \\
			$log$-$Logistic(1,1)$ \end{tabular} & \multicolumn{2}{l|}{\begin{tabular}[c]{@{}l@{}}equal:\\ $C \sim log$-$Normal(\alpha,0.5)$\\ $\alpha=(1.1,0)$ \\ \\ 40\% and 55\%:\\ $C_1,C_2\sim\min \{Exp(0.85),U(0,10)\}$ \\ $C_3,C_4 \sim U(0,10)$ \\ \\ 27\% and 55\%: \\ $C_1 \sim \min \{Exp(0.85),U(0,10)\}$\\
				$C_2 \sim \min \{Exp(0.25),U(0,10)\}$ \\ $C_3\sim U(0,10)$\\ $C_4 \sim log-Normal(1.5,0.5)$ \end{tabular}} \\ \hline
		
		\begin{tabular}[c]{@{}l@{}} Null \\ $K=5$ \end{tabular}                                     &  \begin{tabular}[c]{@{}l@{}}$F_1(t) = F_2(t) = F_3(t) = F_4(t) = F_5(t) =$ \\
			$log$-$Logistic(1,1)$ \end{tabular} & \multicolumn{2}{l|}{\begin{tabular}[c]{@{}l@{}}equal:\\ $C \sim log$-$Normal(\alpha,0.5)$\\ $\alpha=(1.1,0)$ \\ \\ 40\% and 55\%:\\ $C_1,C_2\sim\min \{Exp(0.85),U(0,10)\}$ \\ $C_3,C_4,C_5 \sim U(0,10)$ \\ \\ 27\% and 55\%: \\ $C_1 \sim \min \{Exp(0.85),U(0,10)\}$\\
				$C_2 \sim \min \{Exp(0.25),U(0,10)\}$ \\ $C_3\sim U(0,10)$\\ $C_4 \sim log-Normal(1.5,0.5)$\\ $C_5\sim Exp (1.5)$ \end{tabular}} \\ \hline
		D                           &  \begin{tabular}[c]{@{}l@{}}$F_1(t) = \left\{
			\begin{array}{ll}
			Exp(0.5) & \quad t \le 0.44 \\
			Exp(0.1) & \quad 0.44<t \le 1.05 \\
			Exp(1.5) & \quad 1.05<t \le 1.47 \\
			Exp(1) & \quad t>1.47
			\end{array} 
			\right.$ \\ \\ $ F_2(t)=$\\
			$F_3(t) = \left\{
			\begin{array}{ll}
			Exp(1.5) & \quad t \le 0.38 \\
			Exp(0.1) & \quad 0.38<t \le 1.02 \\
			Exp(0.5) & \quad 1.02<t \le 1.47 \\
			Exp(1) & \quad t>1.47
			\end{array} 
			\right.$ 
		\end{tabular} & \multicolumn{2}{l|}{\begin{tabular}[c]{@{}l@{}}equal:\\ $C \sim U(\alpha,\beta)$\\ $(\alpha,\beta)=(1.1,3), (0.1,2.1)$ \\ \\ 40\% and 55\%:\\ $C_1,C_2\sim\min \{Exp(0.5),U(0.5,3.5)\}$ \\ $C_3\sim U(0.5,3.5)$ \\ \\ 27\% and 55\%: \\ $C_1 \sim \min \{Exp(0.3),U(0.5,3.5)\}$\\
				$C_2 \sim \min \{Exp(0.5),U(0.5,3.5)\}$ \\ $C_3\sim U(0.5,3.5)$ \end{tabular}} \\ \hline
		J-2                           &  \begin{tabular}[c]{@{}l@{}}$F_1(t)= Exp (1)$  \\ \\
			$ F_2(t)=$\\
			$F_3(t)= \left\{
			\begin{array}{ll}
			Exp(1) & \quad t \le 0.1 \\
			Exp(1.7) & \quad 0.1<t \le 0.45 \\
			Exp(0.5) & \quad t>0.45
			\end{array} 
			\right.$ 
		\end{tabular} & \multicolumn{2}{l|}{\begin{tabular}[c]{@{}l@{}}equal:\\ $C \sim Exp(\lambda)$\\ $\lambda=(0.3,1)$ \\ \\ 40\% and 55\%:\\ $C_1,C_2\sim\min \{Exp(0.9),U(0,4)\}$ \\ $C_3 \sim U(0,4)$ \\ \\ 27\% and 55\%: \\ $C_1 \sim \min \{Exp(0.9),U(0,4)\}$\\
				$C_2 \sim  \min \{Exp(0.5),U(0,4)\}$ \\ $C_3\sim U(0,4)$ \end{tabular}} \\ \hline
	\end{longtable}
}

\begin{figure}[!h]
	\caption{Empirical power of the tests  under the null for $K=3,4,5$}
	\centering
	\label{K_null_results}
	\includegraphics[width=1.05\textwidth]{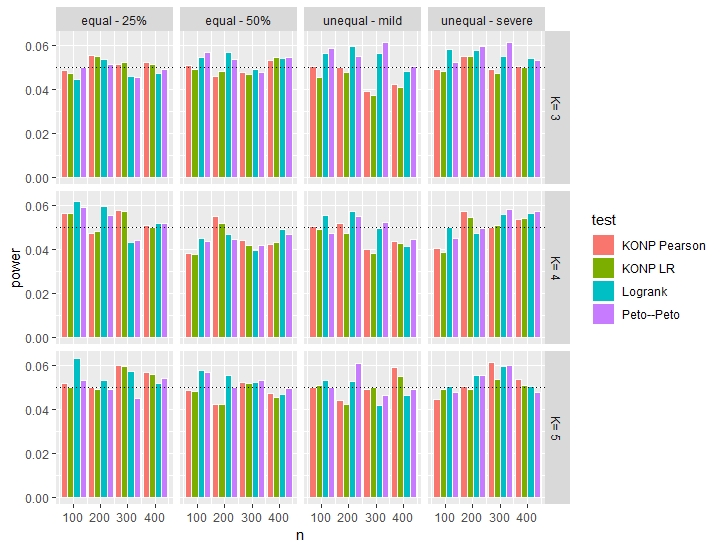}
\end{figure}

\begin{figure}[!h]
	\caption{Empirical power under the alternative for $K=3$}.
	\centering
	\label{K_alternative_results}
	\includegraphics[width=1.15\textwidth]{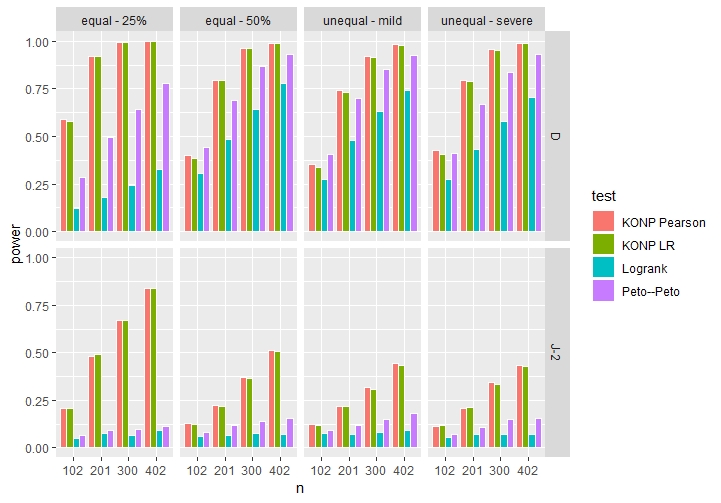}
\end{figure}

\clearpage

\begin{figure}[!h]
	\caption{Empirical power under the null of the 2-sample KONP tests, logrank and the $Cau$ robust test}.
	\centering
	\label{K_alternative_results}
	\includegraphics[width=0.7\textwidth]{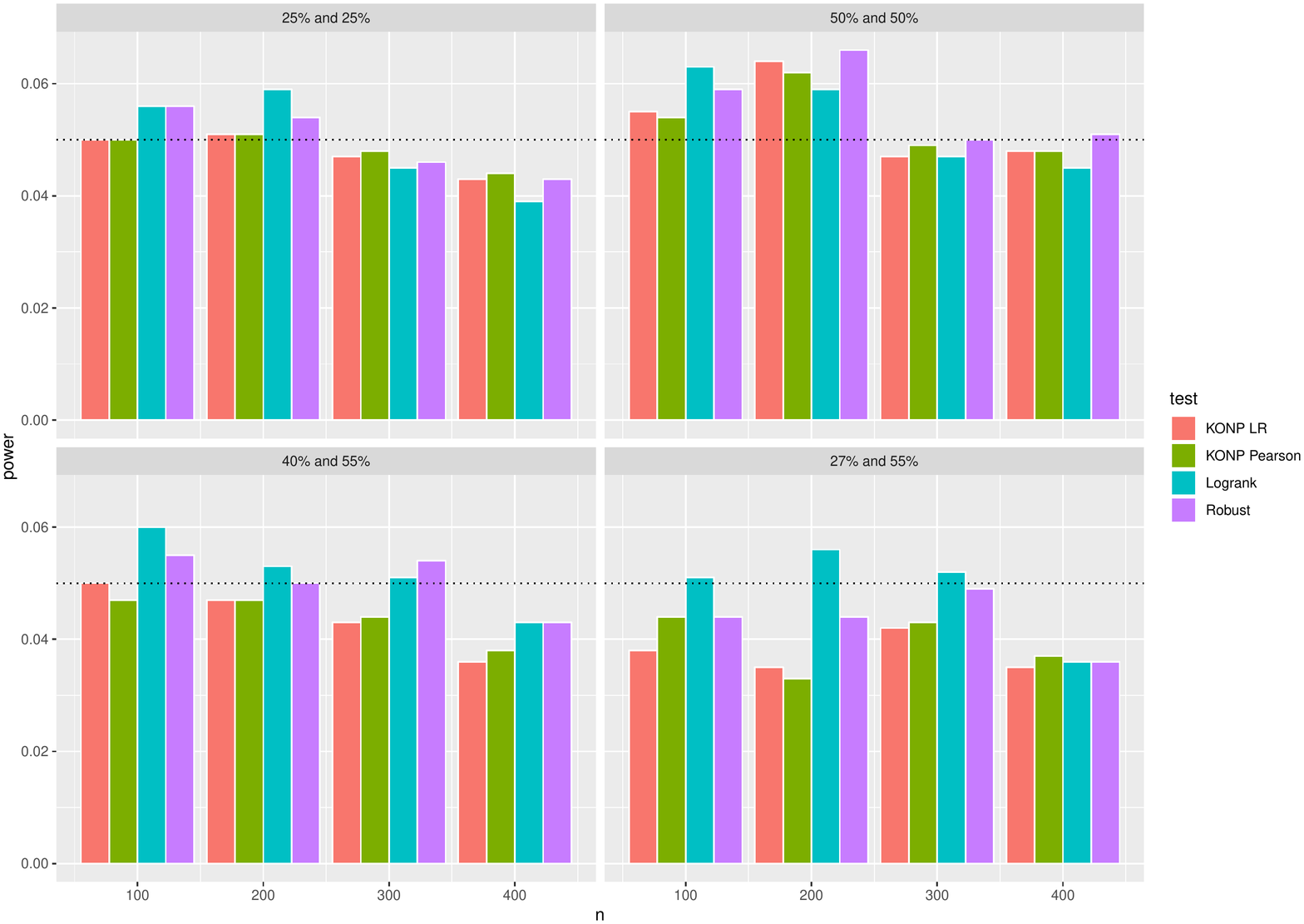}
\end{figure}

\begin{figure}[!h]
	\caption{Empirical power of the 2-sample KONP tests, logrank and the $Cau$ robust test}.
	\centering
	\label{K_alternative_results}
	\includegraphics[width=1.15\textwidth]{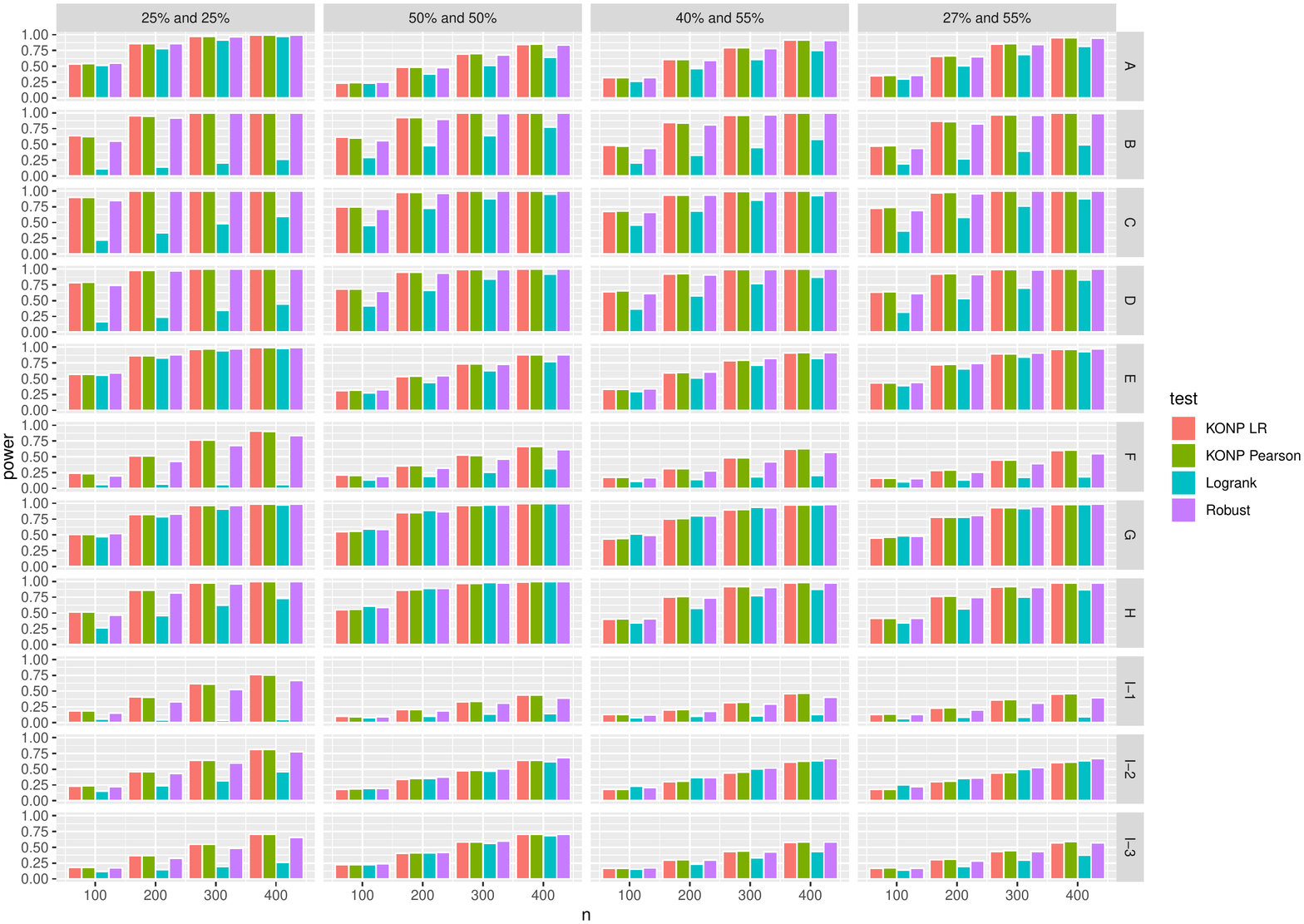}
\end{figure}

\begin{figure}[!h]
	\caption{Empirical power of the 2-sample KONP tests, logrank and the $Cau$ robust test}.
	\centering
	\label{K_alternative_results}
	\includegraphics[width=1.15\textwidth]{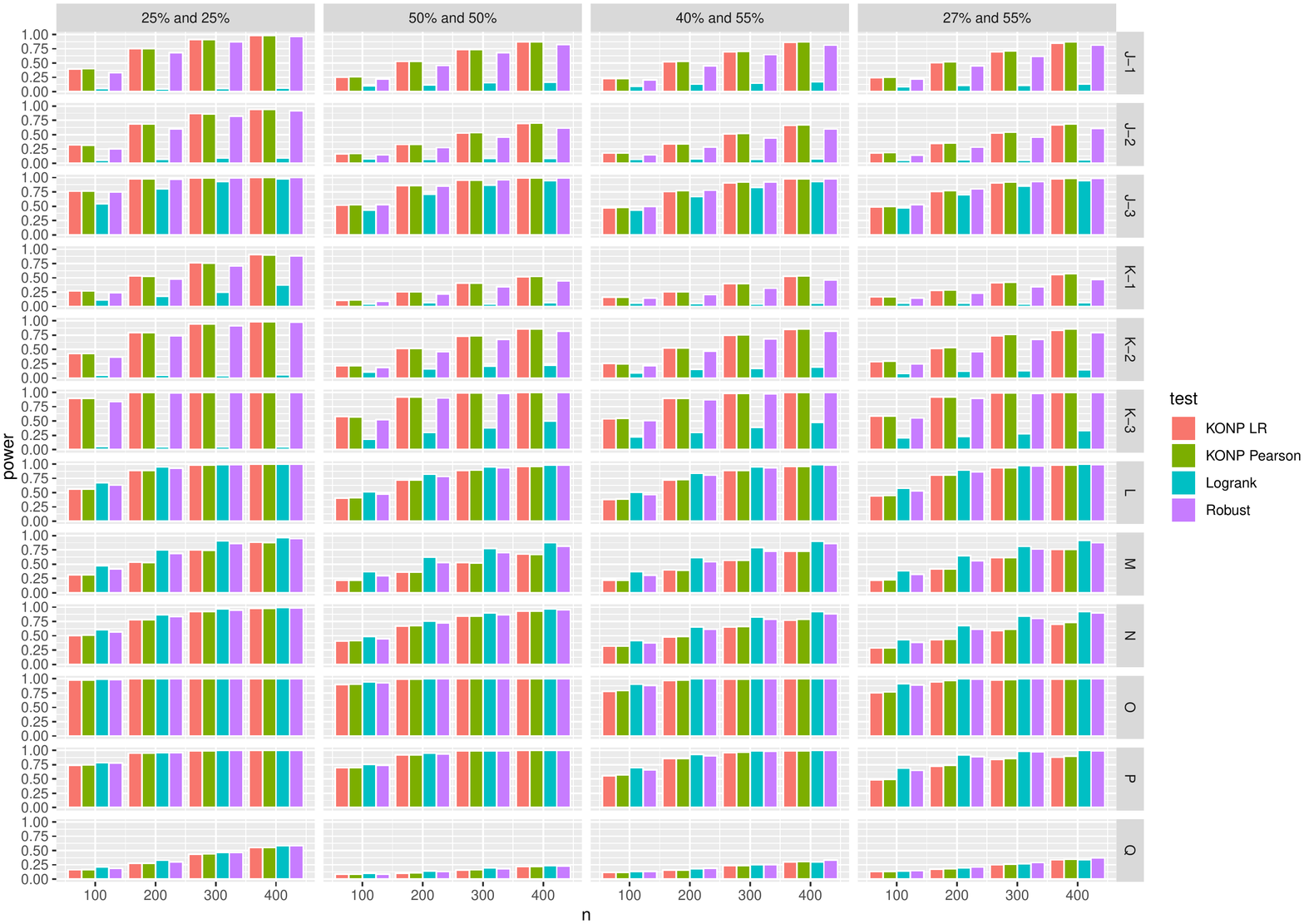}
\end{figure}

\clearpage

\spacingset{1}
\footnotesize
\begin{longtable}[c]{ccccccccccc}
	\caption{Empirical power of all the $2$-sample scenarios: KONP-P and KONP-LR are our two proposed tests; YP-1 and YP-2 are Yang-Prentice tests with group 1 as control and 2 as treatment, and vice verse; YP-Inv is Yang--Prentice invariant test; LR - the logrank test; PP - the Peto--Peto test; and PF - the Pepe--Fleming test. In each line of the table, the highest power \textbf{among the invariant} tests is in bold.
		Evidently, under the non-proportional hazards scenarios (A through K-3) usually KONP-P and KONP-LR are with the highest power. Under proportional hazards (or close to, L through Q) usually the logrank test is with the highest power. }
	\label{2_sample_table} \\ 
	\hline
	n & Scenario & Censoring  & KONP-P & KONP-LR & YP-1 & YP-2 & YP-Inv & LR & PP & PF \\ 
	\hline
	100 & Null & 25\% and 25\%& 0.047 & 0.046 & 0.064 & 0.066 & 0.052 & 0.054 & 0.051 & 0.047 \\ 
	100 & Null & 50\% and 50\% & 0.049 & 0.049 & 0.072 & 0.074 & 0.057 & 0.056 & 0.056 & 0.051 \\ 
	100 & Null & 27\% and 55\% & 0.041 & 0.038 & 0.061 & 0.066 & 0.050 & 0.048 & 0.048 & 0.041 \\ 
	100 & Null & 40\% and 55\% & 0.046 & 0.044 & 0.060 & 0.061 & 0.052 & 0.049 & 0.048 & 0.043 \\ 
	\hline
	100 & A & 25\% and 25\%& 0.492 & 0.492 & 0.513 & {0.556} & \textbf{0.496} & 0.479 & 0.320 & 0.337 \\ 
	100 & A & 50\% and 50\%& \textbf{0.235} & \textbf{0.235} & {0.237} & 0.295 & 0.217 & 0.211 & 0.147 & 0.169 \\
\hline
\end{longtable}

\begin{longtable}[c]{cccccccc}
	\caption{Empirical power of all scenarios with $K=3,4,5$ groups: KONP-P and KONP-LR are our two proposed KONP-Pearson and KONP-likelihood ratio test, respectively; LR and PP are the logrank the Peto--Peto tests}
	\label{K_sample_table} \\ 
	\hline
	K & Scenario & n & Censoring & KONP-P & KONP-LR & LR & PP \\ 
	\hline
	3 & Null &  102 & equal - 25\% & 0.049 & 0.047 & 0.045 & 0.050 \\ 
	
	3 & Null &  102 & equal - 50\% & 0.051 & 0.049 & 0.055 & 0.057 \\ 
	
	3 & Null &  102 & unequal - mild & 0.051 & 0.046 & 0.057 & 0.059 \\ 
	
	3 & Null &  102 & unequal - severe & 0.049 & 0.048 & 0.058 & 0.052 \\ 
	\hline
	3 & D &  102 & equal - 25\% & \textbf{0.587} & 0.581 & 0.122 & 0.286 \\ 
	
	3 & D &  102 & equal - 50\%  & 0.399 & 0.382 & 0.304 & \textbf{0.440} \\
\hline
\end{longtable}

\newpage
\begin{longtable}[c]{ccccccccc}
	\caption{Empirical power of KONP-P and KONP-LR, our two proposed KONP-Pearson and KONP-LR tests, respectively; LR - the logrank test, and the robust test based on the Cauchy-combination test, $Cau$, of Section 3.4 of the main text; {the test of Lee (2007) and the MaxCombo test}. Scenarios L through Q are of proportional hazards or close to it.}\\
	\hline
	$n$ & Scenario & Censoring & KONP-P & KONP-LR & logrank & $Cau$ & {Lee (2007)} & {MaxCombo}\\ 
	\hline
	100 & A & 25\% and 25\% & 0.542 & 0.539 & 0.515 & 0.547 & 0.582 & 0.584 \\ 
	100 & A & 50\% and 50\% & 0.238 & 0.234 & 0.235 & 0.252 & 0.283 & 0.291 \\ 
	100 & A & 40\% and 55\% & 0.323 & 0.322 & 0.262 & 0.320 & 0.333 & 0.329 \\ 
	100 & A & 27\% and 55\% & 0.355 & 0.351 & 0.300 & 0.354 & 0.344 & 0.350 \\ 
	\hline
	200 & A & 25\% and 25\% & 0.858 & 0.860 & 0.782 & 0.856 & 0.867 & 0.878 \\ 
	200 & A & 50\% and 50\% & 0.484 & 0.482 & 0.375 & 0.476 & 0.537 & 0.521 \\
\hline
\end{longtable}

\end{document}